\documentclass[iop, twocolumn]{aastex6}
\usepackage{amsmath}     
\usepackage{wasysym}           
\usepackage{graphicx}
\usepackage{amssymb}
\usepackage{epstopdf}
\usepackage{mathrsfs}
\usepackage{natbib}
\usepackage{color}

\begin{document}
\title{Spherically-symmetric, cold collapse: the exact solutions and a comparison with self-similar solutions}
\shorttitle{cold collapse}
\author{Eric R. Coughlin\altaffilmark{1}}
\shortauthors{\sc{Coughlin}} 
\email{eric$_$coughlin@berkeley.edu}
\affiliation{Astronomy Department and Theoretical Astrophysics Center, University of California, Berkeley, Berkeley 94720, USA}
\altaffiltext{1}{Einstein Fellow}

\begin{abstract}
We present the exact solutions for the collapse of a spherically-symmetric, cold (i.e., pressureless) cloud under its own self-gravity, valid for arbitrary initial density profiles and not restricted to the realm of self-similarity. These solutions exhibit a number of remarkable features, including the self-consistent formation of and subsequent accretion onto a central point mass. A number of specific examples are provided, and we show that Penston's solution of pressureless, self-similar collapse is recovered for polytropic density profiles; importantly, however, we demonstrate that the time over which this solution holds is fleetingly narrow, implying that much of the collapse proceeds non-self-similarly. We show-eps-converted-to.pdf that our solutions can naturally incorporate turbulent pressure support, and we investigate the evolution of overdensities -- potentially generated by such turbulence -- as the collapse proceeds. Finally, we analyze the evolution of the angular velocity and magnetic fields in the limit that their dynamical influence is small, and we recover exact solutions for these quantities. Our results may provide important constraints on numerical models that attempt to elucidate the details of protostellar collapse when the initial conditions are far less idealized.
\end{abstract}
\keywords{galaxies: formation --- gravitation --- magnetohydrodynamics --- methods: analytical --- stars: formation}

\section{Introduction}
Understanding the process by which a cloud of gas collapses under its own self-gravity has obvious physical relevance for a number of problems in astrophysics, including galaxy formation (the ``direct collapse'' model; e.g., \citealt{hae93, kou04, beg06, beg08, joh11, cho15}), the early stages of core-collapse supernovae (e.g., \citealt{lie87, che89, woo05, iwa08}) and long gamma-ray bursts (e.g., \citealt{woo93, mac99, mor07}). However, it is in the field of star formation that this concept has arguably received the most attention. Indeed, the idea of ``protostellar collapse,'' whereby a cloud of gas collapses to form a star (or stars), has been studied for more than half of a century (see \citealt{lar03} and \citealt{mck07} for recent reviews).

Among the first to investigate the problem of protostellar collapse in some detail was \citet{pen69}. In addition to performing some of the first numerical studies, \citet{pen69} also derived one of the first self-similar models for the collapse of a protostellar gas cloud (see also \citealt{lar68, lar69}). Since then many authors have expanded upon the ``Penston-Larson'' self-similar solution (e.g., \citealt{shu77, hun77, whi85, mcl97, mck03, lou06, gon09}){}{, and others have pursued self-similar solutions in the relativistic regime (e.g., \citealt{bar70, cah71, ber85, ori87, lyn88, lem89}}). \citet{pen69} also derived a similarity solution in the limit that the pressure of the gas is neglected (i.e., when the gas is ``cold''), and \citet{lyn88} expanded on that solution. 

Inherent in the assumption of self-similarity is the notion that the initial conditions of the collapsing gas cloud are rapidly forgotten: while the initial cloud may have some inhomogeneities in, e.g., its density profile ($\rho$), as the fluid collapses inwards it is rapidly driven to a simple, self-similar form of the type $\rho \propto r^{-n}$, where $r$ is radial distance from the center of the cloud and $n$ is a number greater than or equal to zero. However, it is not obvious that this state will necessarily be reached in a small fraction of the infall timescale, so that much of the collapse of the protostellar cloud could be non-self-similar even if the asymptotic state is a self-similar one. A relevant question is thus related to the longevity and permanence of the self-similar evolution of a collapsing cloud, even though it may be possible theoretically. 

In this paper we investigate, from a predominantly analytic perspective, the collapse of a spherically-symmetric dust (i.e., a pressureless fluid) cloud. Through a suitable change of variable, in Section \ref{sec:equations} we find solutions to the nonlinear momentum equation and the continuity equation that are valid for \emph{arbitrary} initial density profiles of the cloud, and are thus not restricted to the realm of self-similarity. We also demonstrate, via a proof by contradiction, that these are the \emph{unique} solutions to the problem. Therefore, in our analysis here we solve the coupled, nonlinear, partial differential equations that govern the collapse of a pressureless, spherically-symmetric body under its own self-gravity in their full generality. 

In Section \ref{sec:solutions} we provide a number of examples with specific initial density profiles, including a constant-density cloud, ``nearly-constant-density'' clouds, and polytropic clouds {}{(the polytropic aspect only refers to the initial density profile -- the pressure gradient is not included in the dynamics)}, and we show how those clouds evolve as they collapse; we also consider the case of an infinite, isothermal sphere, which corresponds to a polytropic index of $\gamma = 1$. We demonstrate that, in each of these cases, the center of the cloud self-consistently forms a singularity (the ``star'' for protostellar collapse, or the black hole for the direct collapse model of galaxy formation), and a significant fraction (greater than $\sim90\%$) of the remainder of the cloud accretes onto the central object in a small fraction of an infall time. 

In Section \ref{sec:conditions} we examine the properties of the flow at the time that the singularity forms. We show that, if the initial cloud has a finite central density at the moment collapse begins, then the time taken to form the singularity is $\tau_{coll} = \sqrt{3\pi/(32G\bar\rho)}$, where $\bar\rho$ is the average density of the cloud; thus, gas clouds with the same average density form central singularities at the same time, \emph{independent} of any other properties of their initial density profiles. We also demonstrate agreement with Penston's self-similar solution for polytropic density profiles, and we also verify the findings of \citet{lyn88} for ``flatter'' initial density profiles; however, these self-similar scalings are only manifested over a very small window in time following the formation of the point mass, and thus do not characterize the bulk of the evolution of the gravitational collapse. We calculate the accretion rates at the time the point mass forms, and for density profiles that fall off less than $\rho \propto r^{-3}$ (and thus have infinite mass), we calculate the asymptotic accretion rate onto the central object; for an isothermal sphere, which has $\rho \propto r^{-2}$, we show that the accretion rate asymptotically approaches a constant that is roughly three times the value predicted by \citet{shu77}. We also consider general density profiles that scale as power-laws at large radii and calculate their accretion rates.

In Section \ref{sec:pressure} we show that turbulent pressure support -- which is known to be important in resisting collapse in star forming regions \citep{ful92, mye92} -- can be self-consistently included in our solutions if the turbulent velocity scales with the local infall velocity, which may be actualized in systems that convert gravitational energy into turbulent kinetic energy through a dissipative cascade \citep{mur15}. We also use the generality of our solutions to analyze the temporal evolution of overdensities within the collapsing cloud in Section \ref{sec:overdensities}; accounting for their self-gravity, such overdensities can either be tidally sheared apart or collapse to form infinitely-thin shells, and we derive the condition that differentiates between these two scenarios. 

We discuss and summarize our results in Section \ref{sec:conclusions}, and in Appendices \ref{sec:rotation} and \ref{sec:magnetic} we analyze, respectively, the effects of rotation and magnetic fields on the collapse.

\section{Equations and the general solution}
\label{sec:equations}
\subsection{Equations}
Consider a spherically-symmetric, pressureless cloud that at $t=0$ falls inward from rest due to its own self-gravity. The equations describing the evolution of the cloud are then the radial momentum equation, the continuity equation, and the Poisson equation, which read, respectively,

\begin{equation}
\frac{\partial{v}}{\partial{t}}+v\frac{\partial{v}}{\partial{r}} = -\frac{\partial\phi}{\partial{r}}, \label{rmom}
\end{equation}
\begin{equation}
\frac{\partial\rho}{\partial{t}}+\frac{1}{r^2}\frac{\partial}{\partial{r}}\left(r^2\rho{v}\right) = 0, \label{cont}
\end{equation}
\begin{equation}
\frac{1}{r^2}\frac{\partial}{\partial{r}}\left(r^2\frac{\partial\phi}{\partial{r}}\right) = 4\pi{G}\rho, \label{poisson}
\end{equation}
where $v$ is the radial velocity, $\rho$ is the gas density, and $\phi$ is the gravitational potential.

We can formally integrate Equation \eqref{poisson} to solve for the radial gradient of the gravitational potential:

\begin{equation}
\frac{\partial\phi}{\partial{r}} = \frac{GM(r,t)}{r^2},
\end{equation}
where

\begin{equation}
M(r,t) \equiv 4\pi\int_0^{r}\tilde{r}^2\rho(\tilde{r},t)d\tilde{r} \label{Meq}
\end{equation}
is the mass contained within radius $r$ at time $t$ and $\tilde{r}$ is a dummy variable of integration. With this definition, Equation \eqref{rmom} becomes

\begin{equation}
\frac{\partial{v}}{\partial{t}}+v\frac{\partial{v}}{\partial{r}} = -\frac{GM(r,t)}{r^2}. \label{rmom1}
\end{equation}
We can also multiply equation \eqref{cont} by $4\pi{r}^2$, use the fact that $\partial{M}/\partial{r} = 4\pi{r^2}\rho$, and integrate with respect to $r$ to give

\begin{equation}
\frac{\partial{M}}{\partial{t}}+v\frac{\partial{M}}{\partial{r}} = 0. \label{cont2}
\end{equation}
Equations \eqref{rmom1} and \eqref{cont2} are then the coupled, nonlinear, partial differential equations we need to solve for $v$ and $M$ to determine the evolution of the collapsing dust cloud.

\subsection{General solution}
To solve equations \eqref{rmom1} and \eqref{cont2}, we write the radial velocity as

\begin{equation}
v = -\sqrt{\frac{2GM}{r}}f(\xi), \label{vss}
\end{equation}
where $f$ is a function of the variable

\begin{equation}
\xi = \frac{t\sqrt{2GM}}{r^{3/2}}, \label{xieq}
\end{equation}
which is time normalized by the infall time $\tau_d \sim r^{3/2}/\sqrt{GM}$. It is important to note here that the $M$ appearing in this definition of $\xi$ \emph{is itself a function of $r$ and $t$}, and so we have not actually determined the full functional form of the variable $\xi$ -- it need not be a simple ratio of time to some power of $r$, as would be the case if $M$ were the (constant) mass of an accreting protostar or black hole. 

Inserting Equation \eqref{vss} into Equation \eqref{rmom1} and performing a number of simple but tedious algebraic manipulations gives

\begin{equation}
2f'+f^2+3\xi{f}f'+\frac{r^{3/2}}{M\sqrt{2GM}}\left(f+\xi{f'}\right)\left(\frac{\partial{M}}{\partial{t}}+v\frac{\partial{M}}{\partial{r}}\right) = 1. \label{fss1}
\end{equation}
The last term on the left-hand side of this equation appears to pose a problem as it is not simply a function of the self-similar variable $\xi$. However, we see from Equation \eqref{cont2} that the specific combination of derivatives of $M$ appearing in this term is \emph{uniquely zero}. We therefore find that our ansatz for the radial velocity, Equation \eqref{vss}, results in the following ordinary, nonlinear differential equation for $f$:

\begin{equation}
f' = \frac{1-f^2}{2+3\xi{f}}. \label{fss}
\end{equation}
This equation is exactly the same as that found in \citet{cou16b}, albeit with a sign difference owing to the negative appearing in our definition of the velocity in Equation \eqref{vss}. Because the cloud is assumed to collapse from rest, investigating Equation \eqref{vss} shows that the boundary condition on $f$ is $f(0) = 0$. Thus, the assumption that the radial velocity varies as Equation \eqref{vss} results in a self-consistent solution to the radial momentum equation.

Inserting our self-similar form for $v$ into equation \eqref{cont2} gives our differential equation for $M$:

\begin{equation}
\frac{\partial{M}}{\partial{t}}-\sqrt{\frac{2GM}{r}}f(\xi)\frac{\partial{M}}{\partial{r}} = 0. \label{continuity}
\end{equation}
The continuity equation is thus a nonlinear, partial differential equation, made more complicated because of the fact that $\xi$ depends on $M$ in a way that we have yet to determine. In spite of these difficulties, we find by inspection that a solution is

\begin{equation}
M = g\left(\frac{r}{1-f^2}\right), \label{M0ofxi}
\end{equation}
where $g$ is an arbitrary function that depends only on the variable $r/(1-f^2)$. If we further recall that our boundary condition on the function $f$ is $f(0) = 0$, then the function $g$ is set by the initial conditions of the collapsing cloud, i.e., $g = M_0(r)$, where $M_0(r)$ is the initial distribution of mass contained within radius $r$. We thus have

\begin{equation}
M = M_0\left(\frac{r}{1-f^2}\right). \label{Mofxi}
\end{equation}

One may verify by direct substitution that Equation \eqref{Mofxi} satisfies Equation \eqref{continuity}. However, an alternative, simpler way of doing so is to recognize that the radial momentum equation \eqref{rmom1} may be rewritten as

\begin{equation}
\frac{\partial{B_e}}{\partial{t}}+v\frac{\partial{B_e}}{\partial{r}} = 0,
\end{equation}
where $B_e = v^2/2-GM/r$ is the Bernoulli parameter. It then follows that $M = M(B_e)$ is the general solution to the continuity equation, and using Equation \eqref{vss} for the velocity then yields Equation \eqref{M0ofxi}.

Equation \eqref{Mofxi} is actually an implicit relation for $M$ because of the fact that $\xi$ depends on $M$. However, we can numerically compute the function $f(\xi)$ to a high degree of accuracy, insert the expression into Equation \eqref{Mofxi}, and solve the resulting algebraic relation numerically for $M(r,t)$. This function then determines $\xi$, which in turn gives $v$, closing the solution. Equations \eqref{xieq}, \eqref{fss}, and \eqref{Mofxi} thus determine the solution for the fluid quantities of a cold, collapsing cloud with a spherically-symmetric, but otherwise arbitrary, initial density profile. 

\subsubsection{{}{Lagrangian coordinates}}
{}{The solutions to Equations \eqref{vss}, \eqref{fss}, and \eqref{Mofxi} concern the Eulerian properties of the flow, for which the fluid velocity and mass profile are described in terms of the spatially-fixed radial coordinate $r$ and independent time coordinate $t$. However, an equally-valid approach would be to follow individual fluid elements along their trajectories, thus parameterizing the collapsing cloud in terms of its Lagrangian coordinates $r_i(t)$. Here the subscript $i$ refers to fluid parcel $i$ that was at initial position $r_{0,i}$ at the time that collapse was initiated. }

{}{In a Lagrangian sense, Equation \eqref{Mofxi} represents an integral of motion for the trajectories of the fluid elements. This can be seen by noting that the Lagrangian mass $M[r_i(t)]$ is conserved along flow lines, and hence $M[r_i(t)] = M_0(r_{i,0}) \equiv M_{0,i}$. Using this constraint in Equation \eqref{Mofxi} then shows}

\begin{equation}
\frac{r_i(t)}{1-f(\xi_i)^2} = r_{0,i}, \label{rioft}
\end{equation}
{}{where $\xi_i = \sqrt{2GM_{0,i}}t/r_i(t)^{3/2}$, and this is an algebraic equation that can be numerically solved for $r_i(t)$. This equation can also be derived independently from energy conservation, as the specific energy is $\epsilon_i = v_i^2/2 - GM_i/r_i = GM_i(f(\xi_i)^2-1)/r_i$; since $\epsilon_i$ and $M_i$ are both conserved along flow lines, we again recover Equation \eqref{rioft}. }

\subsection{Uniqueness}
\label{sec:proof}
Equations \eqref{vss}, \eqref{fss}, and \eqref{Mofxi} give solutions for the velocity and density profiles of a collapsing, spherically-symmetric, pressureless cloud, and the only assumption is that the velocity varies as Equation \eqref{vss}. Surprisingly, these solutions are valid for \emph{any} initial density profile that the cloud possesses, but this finding begs the question: are these \emph{the} solutions to the problem, or could there be additional solutions owing to the nonlinearity of the equations?

To answer this question, let us suppose that there exist two other functions, which we will denote $v_1$ and $M_1$, that are solutions to both the radial momentum equation and the continuity equation:

\begin{equation}
\frac{\partial{v_1}}{\partial{t}}+v_1\frac{\partial{v_1}}{\partial{r}} = -\frac{GM_1}{r^2}, \label{radmom1}
\end{equation}
\begin{equation}
\frac{\partial{M_1}}{\partial{t}}+v_1\frac{\partial{M_1}}{\partial{r}} = 0. \label{cont3}
\end{equation}
We further impose the restriction that these functions satisfy the initial conditions $v_1(t=0,r) = 0$ and $M_1(t=0,r) = M_0(r)$. Taking the difference of Equations \eqref{rmom1} and \eqref{radmom1} then gives

\begin{equation}
\frac{\partial}{\partial{t}}\left(v-v_1\right)+v_1\frac{\partial{v_1}}{\partial{r}}+v\frac{\partial{v}}{\partial{r}} = -\frac{G}{r^2}\left(M-M_1\right), \label{diff1}
\end{equation}
while subtracting Equation \eqref{cont3} from Equation \eqref{cont2} yields

\begin{equation}
\frac{\partial}{\partial{t}}\left(M-M_1\right)+v\frac{\partial{M}}{\partial{r}}-v_1\frac{\partial{M_1}}{\partial{r}} = 0. \label{diff2}
\end{equation}
Now, at $t = 0$ the right-hand side of Equation \eqref{diff1} vanishes identically due to the fact that the clouds have the same initial mass profiles, and the last two terms on the left-hand side are also zero because the gas falls from rest. These two points taken together then demonstrate

\begin{equation}
\frac{\partial{v_1}}{\partial{t}}\bigg{|}_{t = 0} = \frac{\partial{v}}{\partial{t}}\bigg{|}_{t = 0}.
\end{equation}
Along the same lines of reasoning, Equation \eqref{diff2} shows that

\begin{equation}
\frac{\partial{M_1}}{\partial{t}}\bigg{|}_{t = 0} = \frac{\partial{M}}{\partial{t}}\bigg{|}_{t = 0}.
\end{equation}
By now differentiating Equations \eqref{diff1} and \eqref{diff2} with respect to time and noting that all cross derivatives (i.e., $\partial^2{v}/\partial{r}\partial{t}$, etc.) appear next to a velocity (either $v$ or $v_1$), it also follows that the second-order time derivatives of $v_1$ and $M_1$ are equal to those of $v$ and $M$. Continuing this process ad-infinitum shows

\begin{equation}
\frac{\partial^{n}v_1}{\partial{t}^n}\bigg{|}_{t = 0} = \frac{\partial^n{v}}{\partial{t}^n}\bigg{|}_{t = 0}, \label{v1deriv}
\end{equation}
\begin{equation}
\frac{\partial^n{M_1}}{\partial{t}^n}\bigg{|}_{t = 0} = \frac{\partial^nM}{\partial{t}^n}\bigg{|}_{t = 0}.
\end{equation}
We may now expand the function $v_1$ in terms of its Maclaurin series:

\begin{equation}
v_1 = \sum_{n=0}^{\infty}\frac{1}{n!}\frac{\partial^nv_1}{\partial{t}^n}\bigg{|}_{t=0}t^{n} = \sum_{n=0}^{\infty}\frac{1}{n!}\frac{\partial^nv}{\partial{t}^n}\bigg{|}_{t=0}t^{n},
\end{equation}
and in the last line we used Equation \eqref{v1deriv} to replace the derivatives of $v_1$ with those of $v$. However, it is apparent that the last expression in this equation is just the Maclaurin series of $v$, which shows that $v_1=v$. Similar reasoning indicates that $M_1=M$, and therefore, if the functions $v$ and $M$ have well-defined series expansions, then \emph{the functions $v$ and $M$ given by Equations \eqref{vss} and \eqref{Mofxi} are the unique, general solutions to the problem of an infalling pressureless gas cloud}. 

In the following section we investigate solutions to those equations for which the initial mass profile, $M_0(r)$, is appropriate to situations in which the cloud density is either exactly constant, approximately constant, or polytropic, and we compare our solutions to those found elsewhere in the literature. {}{Here we are mostly interested in the density and velocity profiles of collapsing clouds, and for these variables the Eulerian approach offers the most straightforward solutions. However, the Lagrangian formalism and the solutions to Equation \eqref{rioft} presents us with a methodology for determining the locations of fluid shells within the cloud, which is useful for conceptualizing the morphological evolution of the collapse. We will therefore also use Equation \eqref{rioft} to follow the infall of individual gas parcels during protostellar collapse.}

\section{specific solutions}
\label{sec:solutions}
\subsection{Constant-density cloud}
\label{sec:const}
\begin{figure*}[htbp] 
   \centering
   \includegraphics[width=0.325\textwidth]{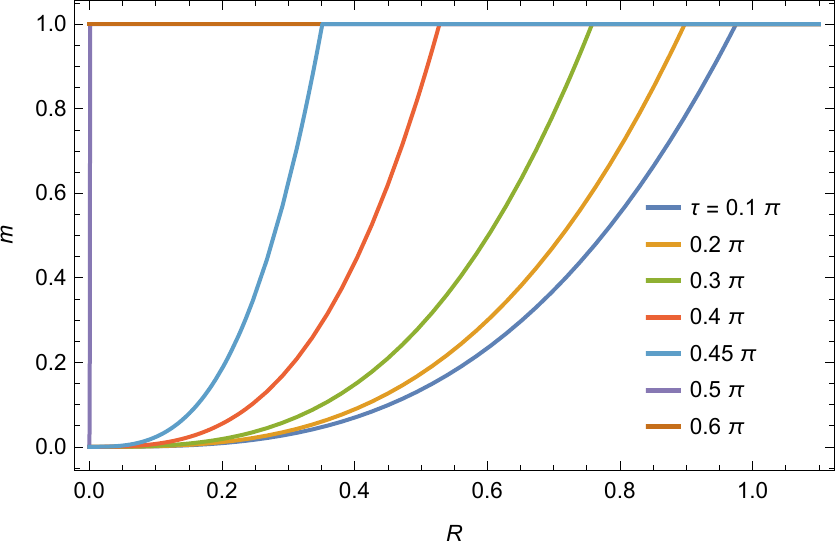} 
   \includegraphics[width=0.325\textwidth]{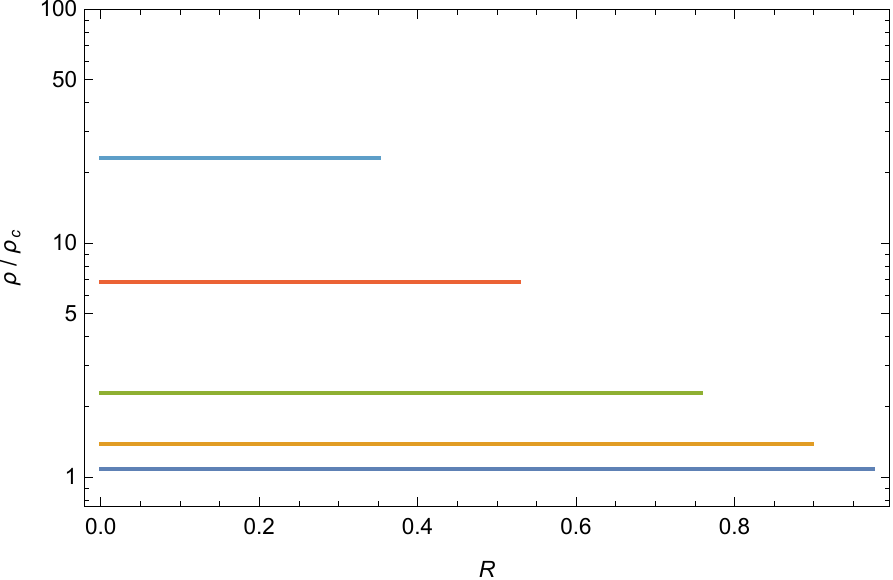}
   \includegraphics[width=0.325\textwidth]{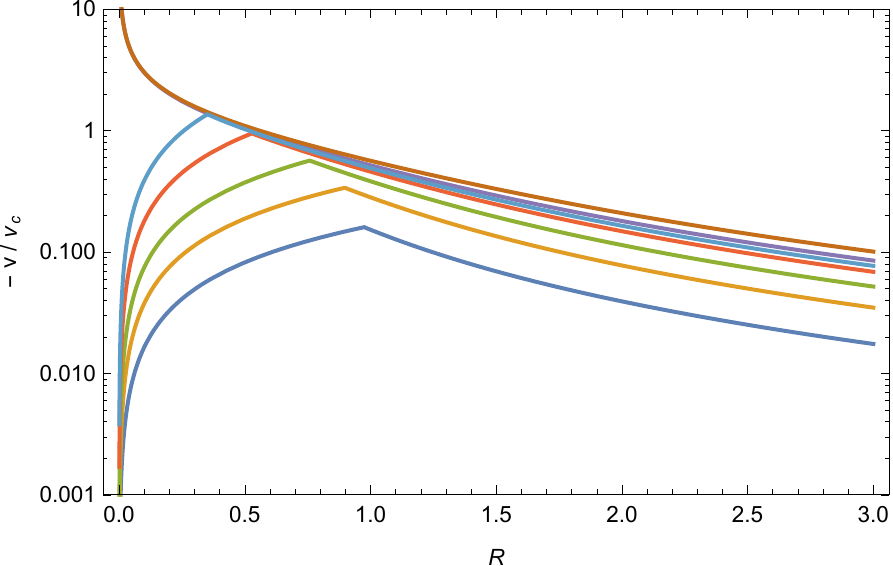}
   \caption{Left: The mass enclosed within radius $R$ as a function of $R$, with the different curves corresponding to the times indicated in the legend. After a time of $\tau \simeq 1.57$, the function $m(r)$ corresponds to a flat line at $m = 1$, showing that the entire self-gravitating cloud has collapsed to a singularity at the origin. Middle: The density as a function of $R$, with the different colors corresponding to the same times as in the left-hand panel. The density goes to zero at the edge of each line, showing that the radius is a decreasing function of time. Right: The velocity as a function of $R$; the cusp in the velocity occurs at the edge of the cloud.}
   \label{fig:mpwofroft}
\end{figure*}

First consider a constant-density cloud of mass $M_C$ and initial radius $R_C$, the initial mass profile for which is

\begin{equation}
M_0(r) = 
\begin{cases}
M_C\left(\frac{r}{R_C}\right)^3\quad\text{for }r \le R_C \\
M_C\qquad\quad\quad\,\,\text{for }r \ge R_C
\end{cases}, \label{mpw}
\end{equation}
Defining $m \equiv M/M_C$, $R \equiv r/R_C$, and $\tau \equiv t/T_C$, where $T_C \equiv R_C^{3/2}/\sqrt{2GM_C}$ scales as the freefall time at the surface of the cloud, it follows from Equation \eqref{Mofxi} that our general solution for the mass contained within $r$ as a function of time is given by the solution to the equation

\begin{equation}
m(R,\tau) = 
\begin{cases}
\left(\frac{R}{1-f^2}\right)^3\quad\text{for }\frac{R}{1-f^2} \le 1  \\ 
\quad1\qquad\quad\,\,\,\text{for }\frac{R}{1-f^2} \ge 1
\end{cases}, \label{mpwoft}
\end{equation}
where, from Equation \eqref{xieq}, we can show that $f$ is a function of $\xi = m^{1/2}R^{-3/2}\tau$. The left-hand panel of Figure \ref{fig:mpwofroft} shows this function, and each curve represents a different dimensionless time as indicated by the legend. This figure demonstrates that, at a time of $\tau = \pi/2$, the radius of the sphere (which is defined as the point at which $m=1$) has collapsed  to the origin, and at every time after that the solution for $m(R)$ is just a flat line exactly at one. We thus find that the entire constant-density sphere collapses to a singularity at a time of $\tau = \pi/2$, which agrees with past findings (e.g., \citealt{pen69}).

The middle panel of Figure \ref{fig:mpwofroft} shows the log of the cloud density, normalized by $\rho_C = 3M_{C}/(4\pi{R_C}^3)$, as a function of $R$ for the same times shown in Figure \ref{fig:mpwofroft}. The edge of each line corresponds to the surface of the cloud (where $m = 1$). Owing to the contracting cloud radius, the value of the density increases as a function of time, growing unbounded at a time of $\tau = \pi/2$.

The right-hand panel of Figure \ref{fig:mpwofroft} illustrates the absolute value of the infall velocity normalized by $v_C = \sqrt{2GM_C/R_C}$ as a function of $R$. The cusp in the solutions coincides with the location of the surface of the cloud. Interior to the cusp, the velocity scales linearly with radius, while outside of the cloud and near its surface the velocity follows $v \propto -1/\sqrt{R}$. However, farther from the surface of the cloud, the velocity transitions to a steeper power-law that is better-matched by $v \propto -1/R^{2}$. This behavior arises because the cloud falls from rest; therefore, gas residing at locations very far from the origin feel a smaller gravitational force and thus take longer to accelerate. This scaling of $v \propto R^{-2}$ can also be demonstrated analytically by noting from $\xi = m^{1/2}R^{-3/2}\tau$ that, when $R \gg 1$, $\xi \ll 1$ and $f \simeq 0$. From Equation \eqref{fss}, we see that $f' \simeq 1/2$, and hence $f \simeq \xi/2 \propto R^{-3/2}$ in this limit. Using this expression for $f$ in Equation \eqref{vss} then shows $v \propto R^{-2}$. It is also straightforward to see that the transition from $v \propto -R^{-1/2}$ to $v \propto -R^{-2}$ occurs when $\xi \simeq 1$, or at a radius of $R \simeq \tau^{2/3}$.

{}{It may seem odd that we are analyzing the velocity profile at points exterior to the surface of the collapsing cloud in Figure \ref{fig:mpwofroft}, as there is no matter in these regions. However, one may consider the velocity profile exterior to the cloud surface as characteristic of more rarefied material, the density of which does not contribute significantly to the gravitational field itself. This situation may be appropriate to collapsing protostellar cores within larger ``clumps,'' (both of which reside within a giant molecular cloud; e.g., \citealt{lar03}) where the density of the latter is not important for the dynamical collapse of the former. However, one may still trace the motion of the gas within the clump by using the velocity profile exterior to the core, which could conceivably be modeled as a constant-density, collapsing sphere.}

\begin{figure}[htbp] 
   \centering
   \includegraphics[width=0.47\textwidth]{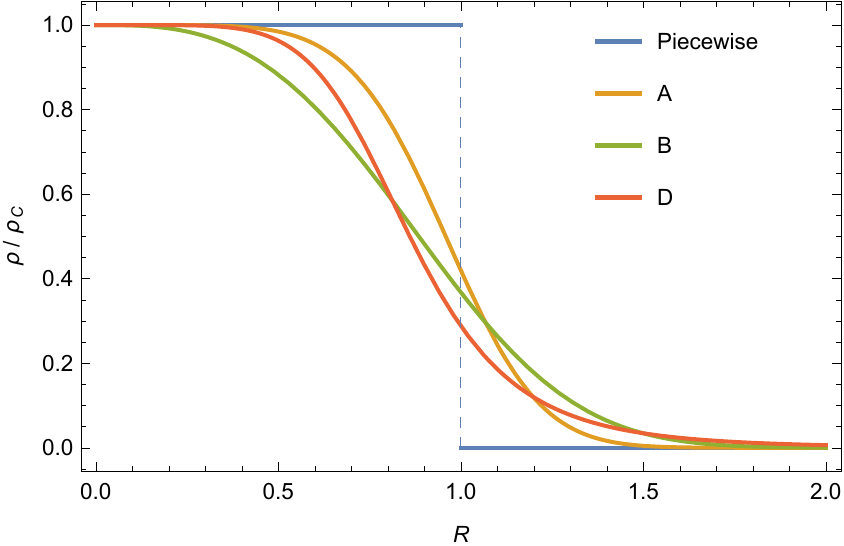} 
   \caption{The initial cloud densities for the initial mass profiles given by $m_{0,A}$ (yellow, dashed curve), $m_{0,B}$ (green, dot-dashed curve), and $m_{0,D}$ (red, dotted curve), and the piecewise solution for a constant-density sphere (equation \ref{mpw}; blue, solid curve). This plot shows how these initial mass profiles approximate the true, constant-density cloud.}
   \label{fig:mplots}
\end{figure}

\subsection{Clouds with near-constant density}
\label{sec:appconst}
A constant-density cloud with a sharp, outer radius is a simple, well-known example, and the results we derived above are in agreement with past investigations (e.g., \citealt{spi68, sim16}). However, the assumption of an infinitely-sharp outer boundary is not very physical, and an interesting question is how sensitive the infall of the cloud is to any deviations from the constant-density assumption.

\begin{figure*}[t!] 
   \centering
   \includegraphics[width=0.325\textwidth]{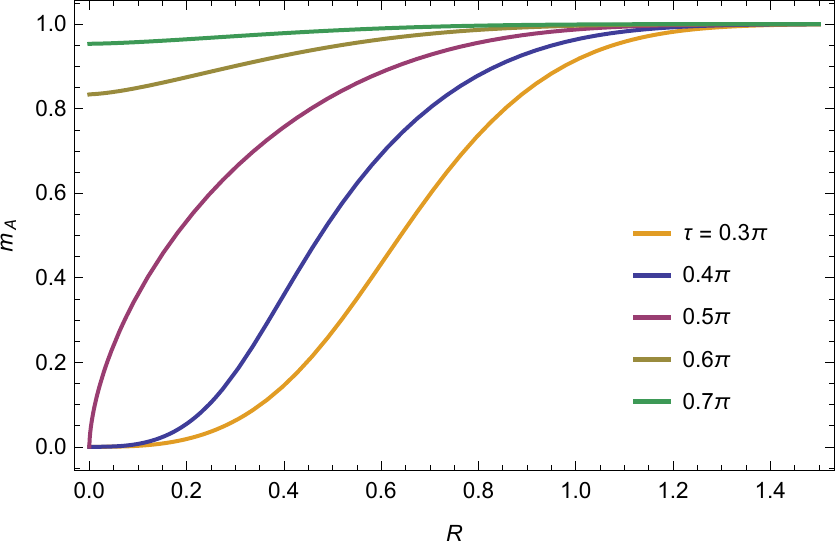} 
   \includegraphics[width=0.33\textwidth]{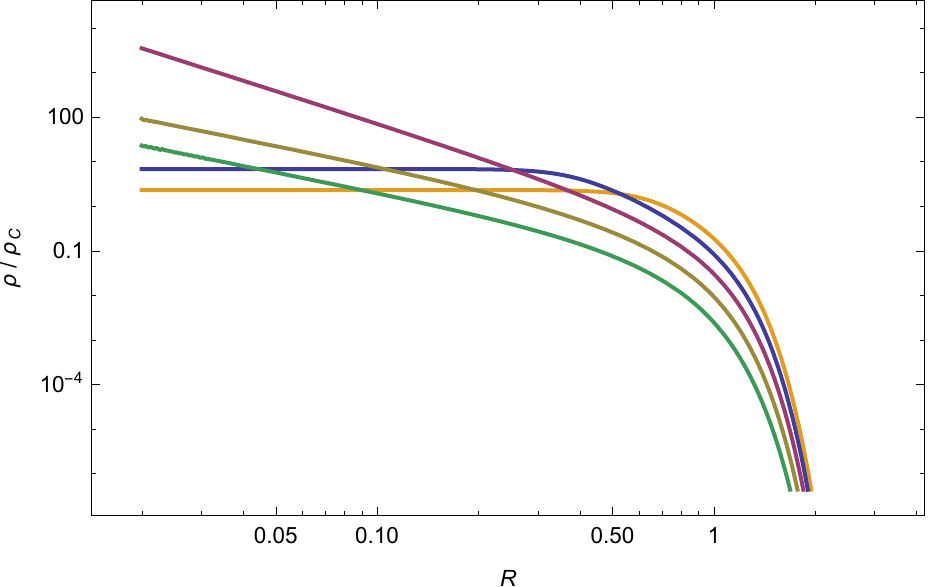} 
   \includegraphics[width=0.33\textwidth]{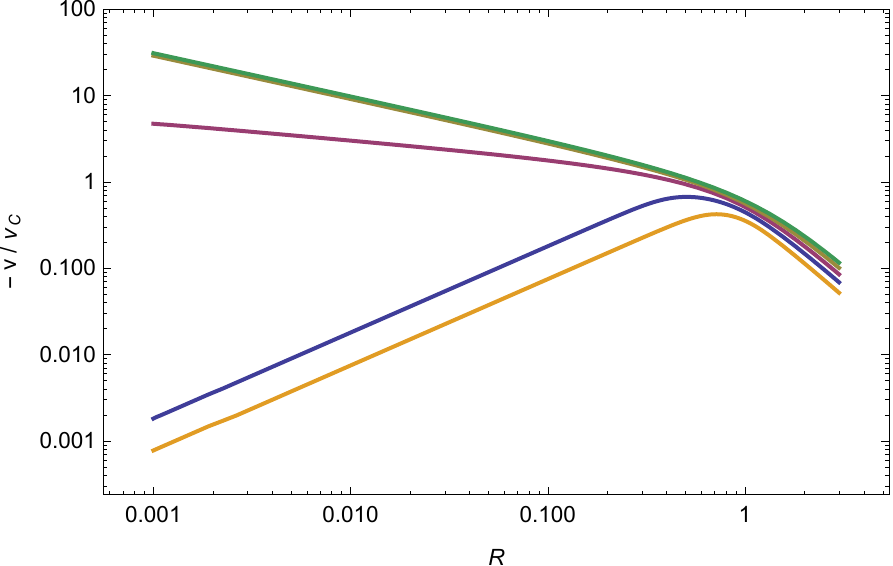} 
   \caption{The mass enclosed within radius $R$ (left panel), the normalized density on a log-log scale (middle panel), and the normalized infall velocity on a log-log scale (right panel) for case $A$ (Equation \ref{m0A}). The different curves correspond to different times, those times given by the legend in the left panel. The left-hand panel shows that, at a time of $\tau = \pi/2$, the center of the cloud collapses to form a singularity. However, unlike the constant-density cloud, only a vanishingly small fraction of the cloud forms the initial singularity, and it takes a finite amount of time to accrete the surrounding gas. We find that, at the time the point mass forms, the density scales as $\rho \propto R^{-2.4}$ in regions near the singularity (i.e., for $R \ll 1$), while the infall velocity follows $v \propto R^{-0.2}$.}
   \label{fig:Aplots}
\end{figure*}

\begin{figure*}[t!] 
   \centering
   \includegraphics[width=0.325\textwidth]{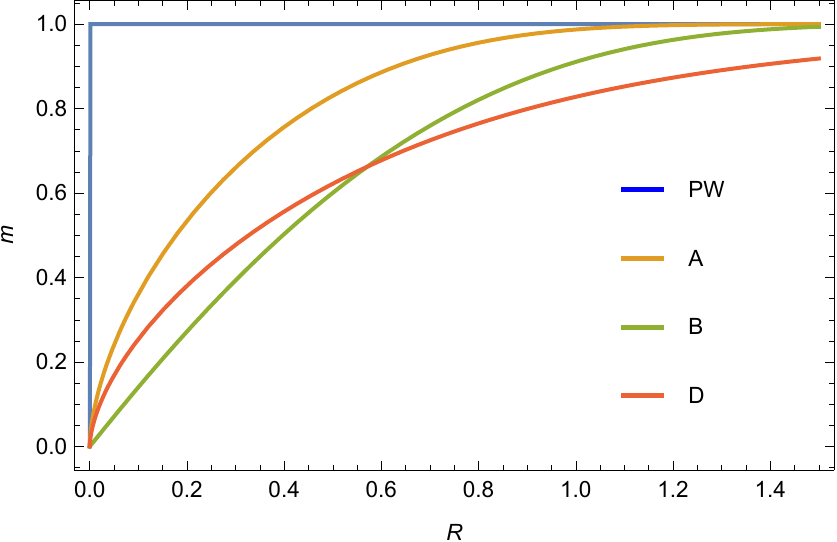} 
   \includegraphics[width=0.33\textwidth]{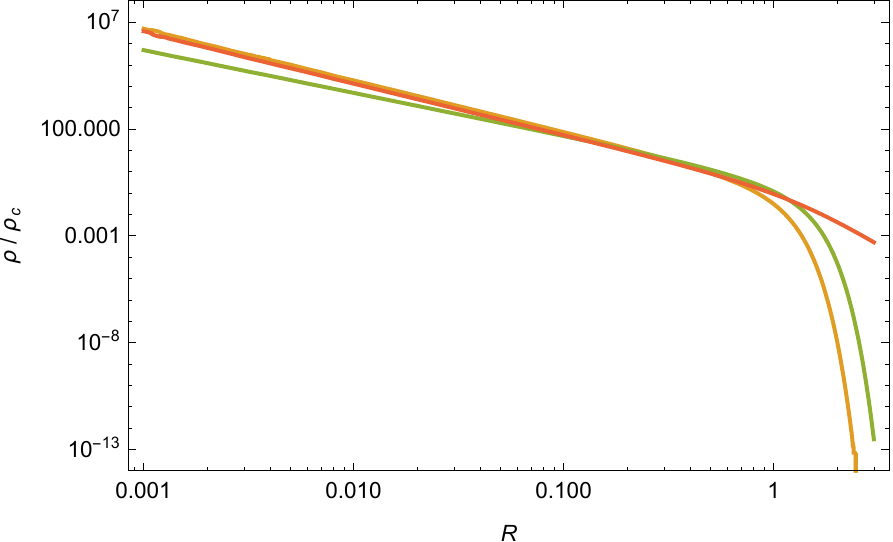} 
   \includegraphics[width=0.33\textwidth]{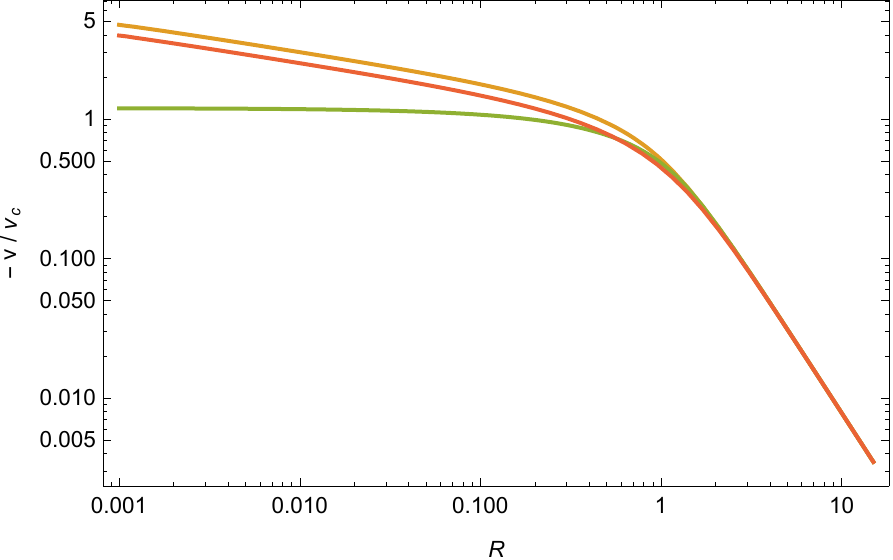} 
   \caption{The mass enclosed within radius $R$ (left panel), the normalized density on a log-log scale (middle panel), and the normalized infall velocity on a log-log scale (right panel) at $\tau = \pi/2$ -- when the singularity first forms at the center of the cloud -- for the constant-density cloud (blue curve), case A (orange, dashed curves), case B (green, dot-dashed curves), and case D (red, dotted curves). Interestingly, these panels demonstrate that cases A and D have very similar density and velocity profiles at this time (being $\rho \propto R^{-2.41}$ and $v \propto R^{-0.23}$), while the initial mass profile appropriate to case B yields $\rho \propto R^{-2.00}$ and $v \propto const.$.}
   \label{fig:crits}
\end{figure*}

We thus consider three initial mass profiles that are appropriate to clouds with ``approximately constant'' densities. In particular, we will let

\begin{equation}
m_{0,A} = \tanh\left(R^3\right), \label{m0A}
\end{equation}
\begin{equation}
m_{0,B} = 1-e^{-R^3}, \label{m0B}
\end{equation}
\begin{equation}
m_{0,D} = \frac{2}{\pi}\arctan\left(\frac{\pi}{2}R^3\right), \label{m0D}
\end{equation}
where, as was done in the previous subsection, $m \equiv M/M_C$ and $R \equiv r/R_C$, with $M_C$ and $R_C$ being the respective cloud mass and radius. Note that, while the value of $R_C$ for the constant-density cloud defined the location at which the density equaled exactly zero, in these instances $R_C$ should be interpreted as a scale length over which the density decreases appreciably (in all cases, $R=1$ corresponds approximately as the $e$-folding radius of the density). Figure \ref{fig:mplots} shows how these functions approximate the true, constant-density cloud (Equation \ref{mpw}).

With Equations \eqref{m0A} -- \eqref{m0D}, we can follow precisely the same procedure as we did in the previous subsection: solve equation \eqref{Mofxi} for $m_A(R,\tau)$, $m_B(R,\tau)$, and $m_D(R,\tau)$, use those solutions to determine the functional form of $\xi$, and use those two combined solutions to evaluate the cloud density $\rho$ and the infall velocity $v$.

Figure \ref{fig:Aplots} shows, from left to right, the mass contained within $R$, the density as a function of $R$, and the infall velocity as a function of $R$ for the initial mass profile given by case $A$ (i.e., Equation \ref{m0A}). The different curves correspond to different times since the onset of collapse, as given in the legend of the left-hand panel of this figure. We see that, at a time of $\tau = \pi/2$, the center of the cloud forms a point mass (i.e., the mass at $R=0$ transitions from being zero to a finite value, which is apparent from the left-hand panel of Figure \ref{fig:Aplots}); however, unlike the constant-density case, the entire cloud does not instantaneously collapse to the origin. Instead, only a vanishingly small amount of material forms the initial singularity, with more than 95\% of the remaining cloud being accreted onto the point mass in roughly half an infall time. 

We find that, at the time that the singularity forms, the density follows the power-law $\rho \propto R^{-2.41}$ in regions close to the point mass, while the infall velocity scales as $v \propto R^{-0.23}$. Interestingly, these deviate substantially from the predictions made by Penston's self-similar solution, those predictions being $\rho \propto R^{-12/7} = R^{-1.71}$ and $v \propto R^{1/7} = R^{0.14}$. We will return to a discussion of why this is in Section \ref{sec:conditions}. For times after the singularity forms, the density varies approximately as $\rho \propto R^{-1.5}t^{-9.4}$ for $R \lesssim 1$, where the very steep temporal power-law follows from the fact that the point mass rapidly accretes the remainder of the cloud. On the other hand, the velocity profile very quickly approaches $v/v_C \simeq -R^{-1/2}$ for $R \lesssim 1$, showing that the solution has settled into a configuration that is dominated by the gravitational field of the central mass. As for the constant-density case, the velocity falls off as $v \propto -R^{-2}$ for $R \gtrsim 1$.

For initial mass profiles B and D, we also find that $\tau = \pi/2$ marks the time at which a singularity forms at the center of the collapsing cloud. Figure \ref{fig:crits} shows the mass contained within $R$ (left panel), the normalized density as a function of $R$ (middle panel), and the normalized infall velocity as a function of $R$ (right panel) for the constant-density cloud and cases A, B, and D at a time of $\tau = \pi/2$. Interestingly, we see that the density and velocity power-laws at this time are not identical for these three cases; in particular, cases A and D satisfy $\rho \propto R^{-2.41}$ and $v \propto -R^{-0.23}$, while case B gives $\rho \propto R^{-2.00}$ and $v \propto const$. Perhaps most surprising about this result is that the functional dependences of cases $A$ and $B$ are \emph{most similar initially} (compare Equations \ref{m0A} and \ref{m0B} and recall the form for the hyperbolic tangent), and yet $A$ and $D$ result in identical scalings near the point mass at the time the core collapses.

\begin{figure}[htbp] 
   \centering
   \includegraphics[width=0.47\textwidth]{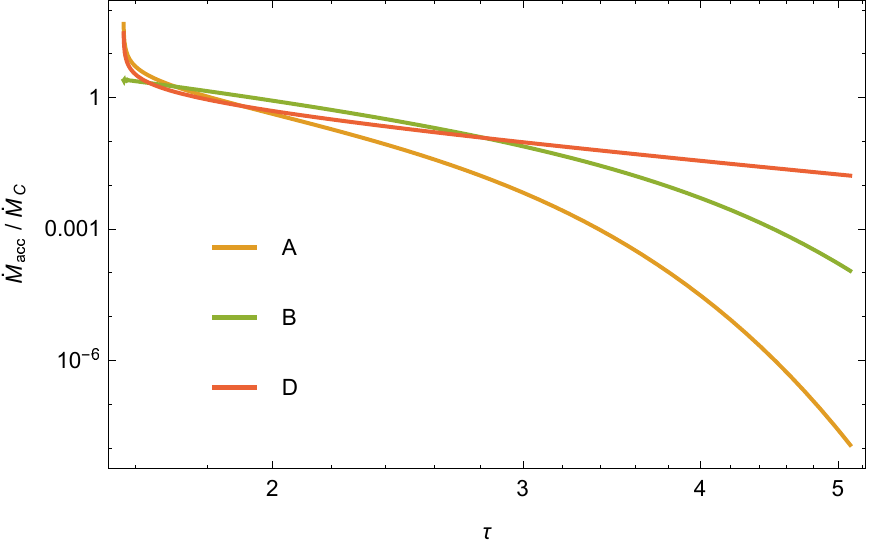} 
   \caption{The accretion rate onto the point mass in units of $\dot{M}_C \equiv M_C/T_C = \sqrt{2G}(M_C/R_C)^{3/2}$ as a function of time, where the different curves represent cases $A$, $B$, and $D$, as indicated by the legend.}
   \label{fig:mdots}
\end{figure}

Figure \ref{fig:mdots} shows the accretion rates as functions of time (on a log-log scale) for cases $A$, $B$, and $D$, with the accretion rates measured in units of $\dot{M}_C \equiv M_C / T_C = \sqrt{2G}M_C^{3/2}/R_C^{3/2}$. This plot demonstrates that the accretion rate falls off as a very steep function of time, which arises from the fact that there is only a finite amount of material contained in the initial cloud and, at the time that the singularity forms, much of that material is moving inward at a significant fraction of the freefall speed. We also see that cases $A$ and $D$ exhibit ``cuspy'' accretion rates as $\tau \rightarrow \pi/2$, with the magnitude of the accretion rate approaching an infinitely large value when the singularity initially forms. On the other hand, the accretion rate for $B$ levels off to a power-law, and we find that the scaling is approximately $\dot{M}_{B} \propto \tau^{-4}$. 

\begin{figure*}[htbp] 
   \centering
   \includegraphics[width=0.325\textwidth]{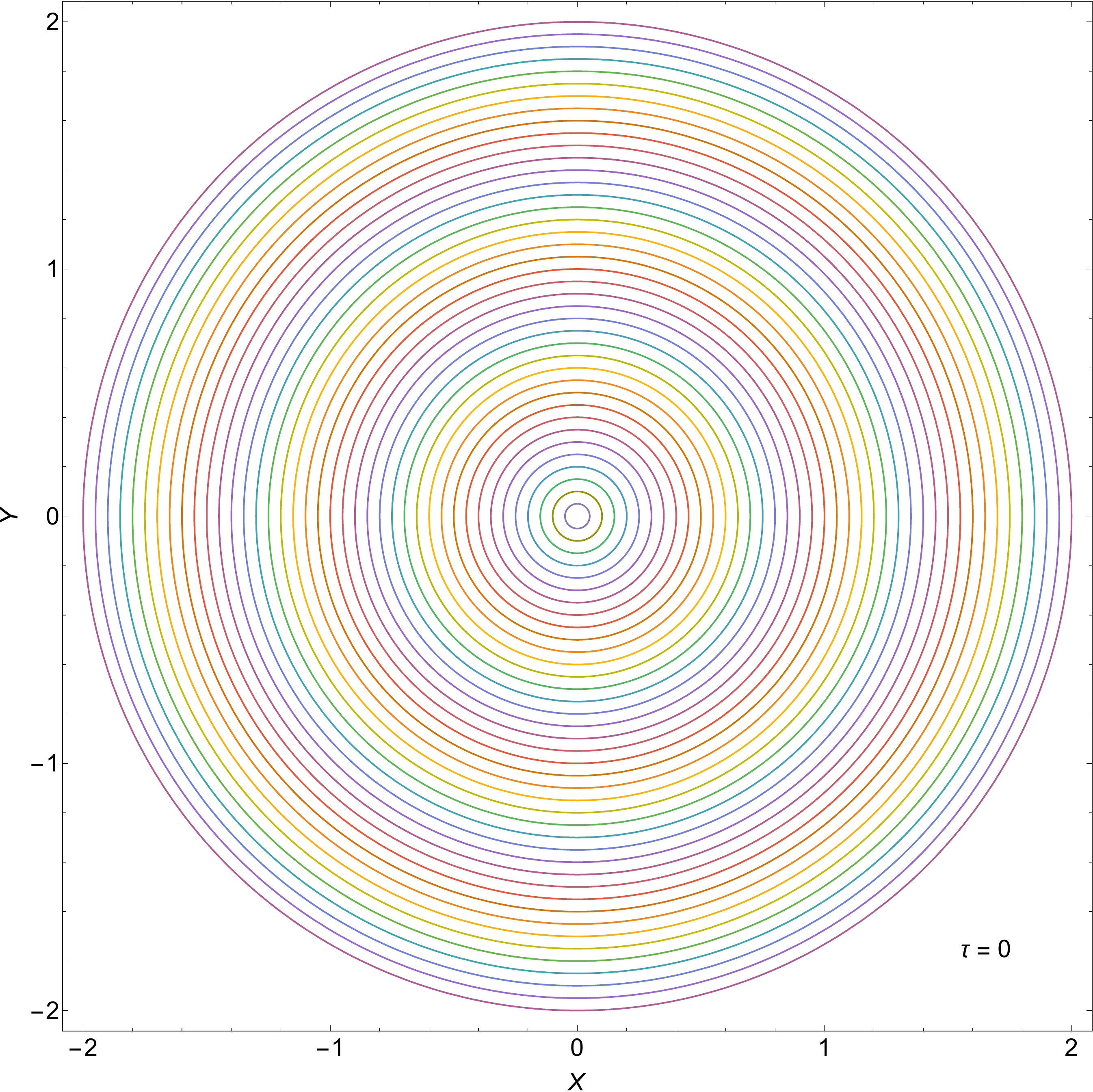} 
   \includegraphics[width=0.325\textwidth]{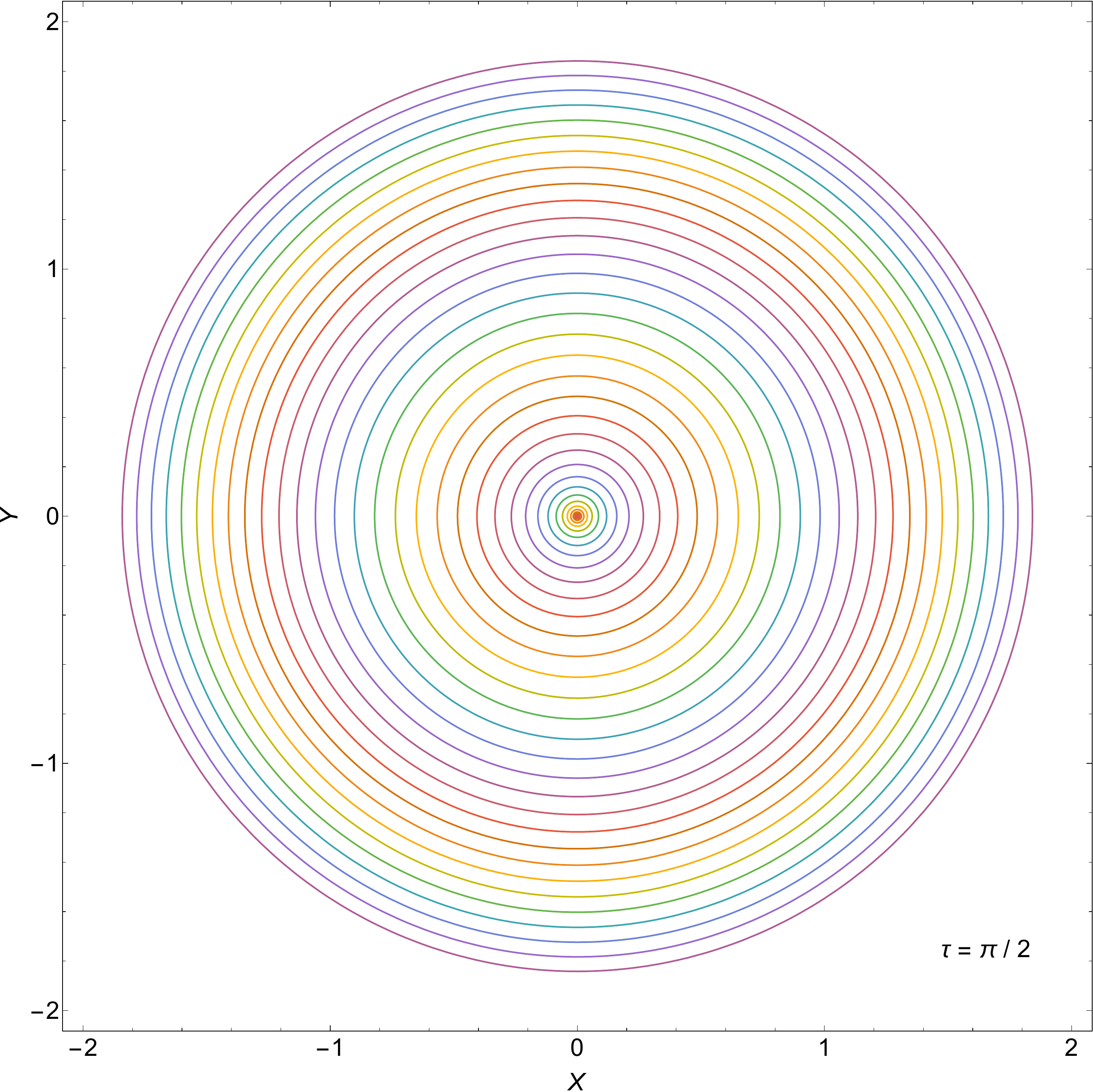}
   \includegraphics[width=0.325\textwidth]{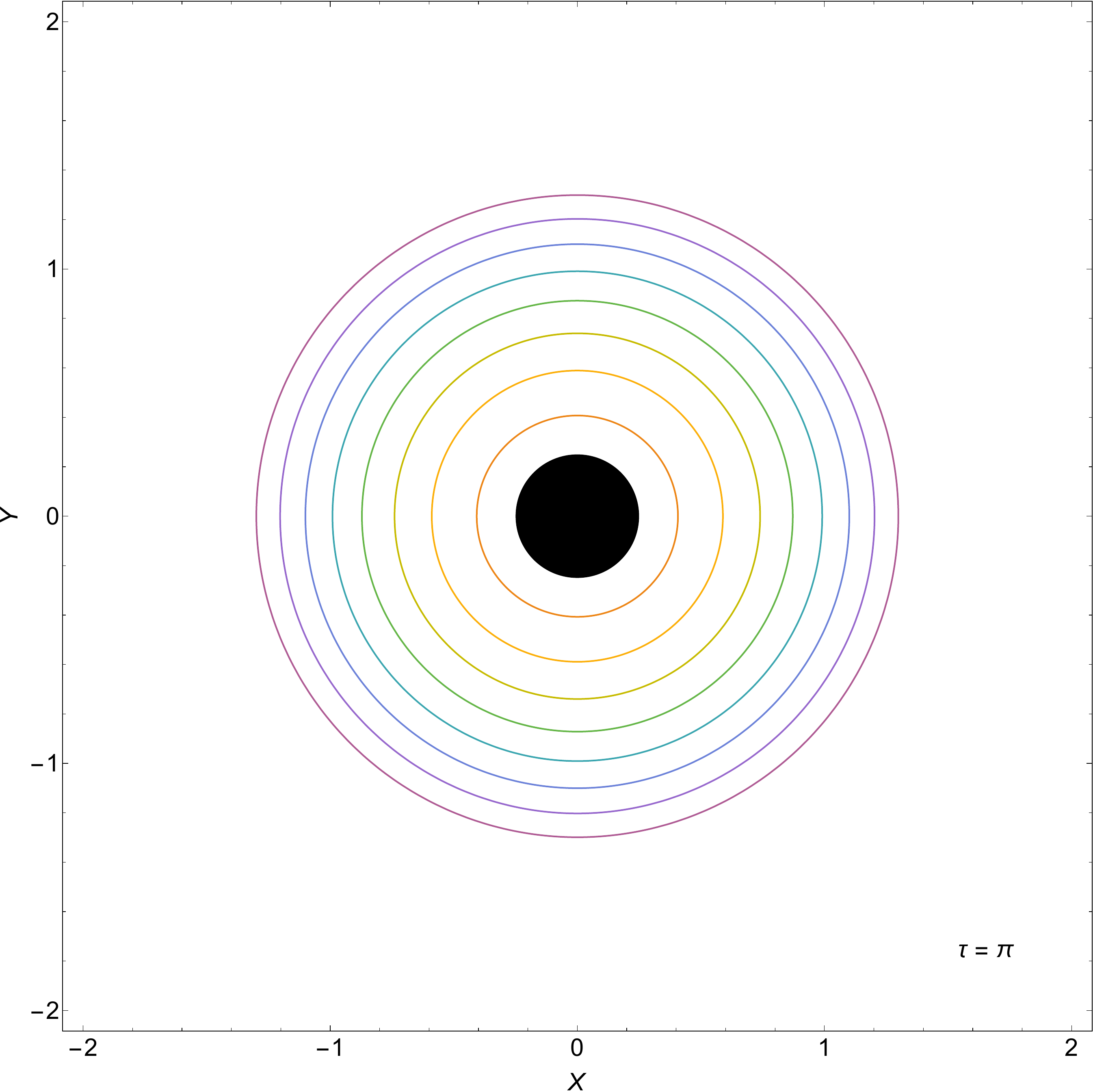}
   \caption{{}{The Lagrangian positions of fluid shells within the collapsing cloud at $\tau = 0$ (left panel), $\tau = \pi/2$ (middle panel), and $\tau = \pi$ (right panel); the different colors merely distinguish between different shells, each of which starts at a different initial radius (shown in the left-hand panel). The black circle denotes the fact that a point mass has formed, and the size of the circle is drawn in proportion to the amount of mass that has been accreted. An animation of the collapse can be found \href{http://w.astro.berkeley.edu/~eric_coughlin/movies.html}{here}.}}
   \label{fig:infall}
\end{figure*}

{}{Figure \ref{fig:infall} shows the Lagrangian evolution of the fluid shells within the collapsing cloud for case A, obtained by solving Equation \eqref{rioft} for a number of different $r_{i,0}$; the different colors merely distinguish between different shells. The left-hand panel shows the configuration at $\tau = 0$ (and therefore gives the initial radii $r_{i,0}$), the middle panel depicts the distribution of material at $\tau = \pi/2$ (when the point mass forms), and the right-hand panel shows, at a time of $\tau = \pi$, how material continues to accrete onto the point mass that has formed (the point mass is shown by the black circle, the radius of which is drawn in proportion to the amount of material it has accreted).}

{}{These representations are useful for visualizing the way in which the density responds to the collapse, which is characterized by the change in distance between neighboring shells. Thus, at a time of $\tau = \pi/2$, we see that the innermost shells have become immensely squeezed in the central region of the cloud, which ultimately leads to the formation of the singularity. At later times, the tidal shear of the point mass causes neighboring shells to accelerate away from one another, resulting in a decrease in the density that is independent of the fact that there is less mass remaining in the cloud at later times. A movie showing the evolution of the collapse can be found \href{http://w.astro.berkeley.edu/~eric_coughlin/movies.html}{here}.}

\subsection{Polytropes}
\label{sec:polytropes}
A physically-motivated, well-studied (see, e.g., \citealt{pen69b, fat04, lou14}) initial mass profile {}{(i.e., before the loss of pressure support)} of a collapsing cloud is that of a polytrope -- one in which the pressure scales as $p = K\rho^{\gamma}$, where $p$ is the pressure, $\gamma$ is the polytropic index, and $K$ is a constant that scales with the specific entropy of the gas. With this form for the pressure, the equation of hydrostatic equilibrium combined with the Poisson equation gives the Lane-Emden equation (e.g., \citealt{han04}):

\begin{equation}
\frac{1}{R^2}\frac{d}{dR}\left(R^2\frac{d\theta}{dR}\right) = -\theta^{\frac{1}{1-\gamma}}, \label{LE}
\end{equation}
where $R = r/\alpha$ is the normalized radius of the polytropic sphere and $\theta = \rho/\rho_0$ is the normalized density, $\rho_0$ being the density at the center of the polytrope and $\alpha^2 = K\gamma\rho_0^{\gamma-2}/[{4\pi{G}(\gamma-1)]}$ the scale length. The boundary conditions are then $\theta(0)=1$ and $\theta'(0) = 0$, while the mass contained within radius $R$ is

\begin{equation}
M(R) = 3M_0\int_0^{R}\tilde{R}^2\theta^\frac{1}{\gamma-1}d\tilde{R},
\end{equation}
where

\begin{equation}
M_{0} = \frac{4\pi}{3}\alpha^3\rho_0
\end{equation}
is the scale mass. {}{Solutions to this equation then describe the initial distribution of gas within the cloud before collapse ensues.}

Due to the nonlinear nature of the Lane-Emden equation, polytropes with $\gamma > 6/5$ have finite radii and masses. A common approach in comparing these polytropes is to then normalize the total enclosed mass and total radius to the same value. However, because it is consistent with what we have done in previous sections, we will instead opt to consider polytropes with the same \emph{scale mass} $M_0$ and the same \emph{scale radius} $\alpha$. This method is also reasonable from a physical standpoint, as any realistic cloud of gas does not have an infinitely sharp outer boundary, even though the scale lengths and masses remain well-defined. 

We will also consider isothermal clouds, which have $\gamma = 1$; for this case we set $\rho = \rho_0\theta$, $\alpha^2 = K/(4\pi{G}\rho_0)$, and the value of $K$ is $K = kT/\mu$, where $k$ is Boltzmann's constant, $T$ is the gas temperature, and $\mu$ is the mean molecular weight, and the Lane-Emden for this case becomes

\begin{equation}
\frac{1}{R^2}\frac{d}{dR}\left(R^2\frac{d\ln\theta}{dR}\right) = -\theta. \label{isoth}
\end{equation} 
As we noted above, only equations of state with $\gamma > 6/5$ have finite radii. Furthermore, for an isothermal sphere the density asymptotically falls off as $\rho \propto r^{-2}$, meaning that the mass of such a sphere is infinite. Nevertheless, this offers no computational difficulty for the equations, and we allow the mass in this case to be infinite. 

\begin{figure}[htbp] 
   \centering
   \includegraphics[width=0.47\textwidth]{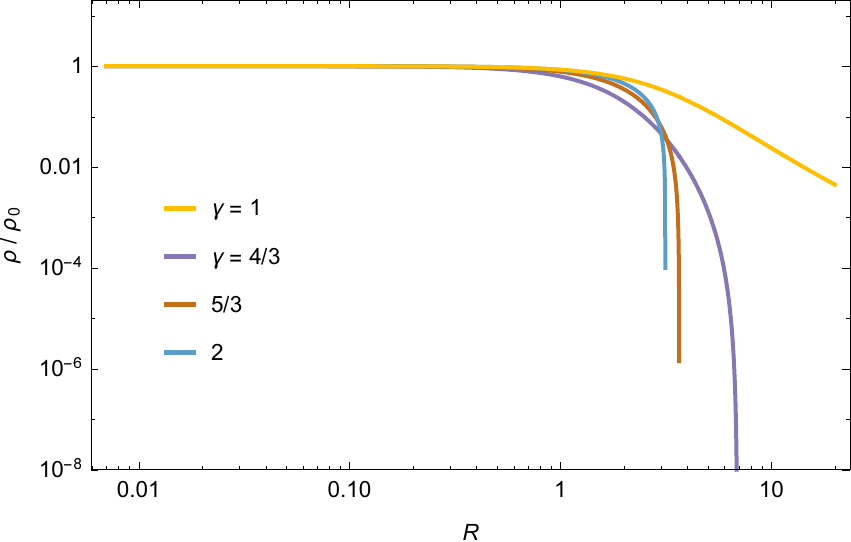} 
   \caption{A log-log plot of the initial density profiles for $\gamma = 4/3$, 5/3, and 2, as indicated by the legend. }
   \label{fig:polytrope_init}
\end{figure}

Figure \ref{fig:polytrope_init} shows the initial density profiles of polytropes with $\gamma = 1$, $4/3$, 5/3, and 2 (the higher densities for lower $\gamma$'s arise from the fact that the masses and radii of the clouds are the same). Physically, $\gamma = 4/3$ corresponds to clouds that were previously pressure supported by relativistic particles (photons, relativistic electrons, or cosmic rays) or a disorganized magnetic field \citep{hei00}, $\gamma = 5/3$ holds if non-relativistic gas pressure supported the cloud prior to collapse, and $\gamma = 2$ if the cloud was supported by an ordered magnetic field. The isothermal case has been studied extensively in the literature, and physically corresponds to the case where the gas temperature is balanced by heating from background radiation (or cosmic rays) and line or dust cooling \citep{lar73}. It is thought that the collapse of a cloud will, in general, sample a number of different adiabatic indices throughout its contraction \citep{bat98}, and we consider these different cases to highlight the effects of varying this quantity. {}{Note that the Lane-Emden equation is only used to \emph{initialize} the mass profile of the cloud, and, as above, the pressure gradient is ignored in the ensuing gravitational collapse. As in Sections \ref{sec:const} and \ref{sec:appconst}, we are therefore solving Equations \eqref{vss}, \eqref{fss}, and \eqref{Mofxi}, with the difference being in the initial density profile of the collapsing cloud.}

\begin{figure*}[htbp] 
   \centering
   \includegraphics[width=0.325\textwidth]{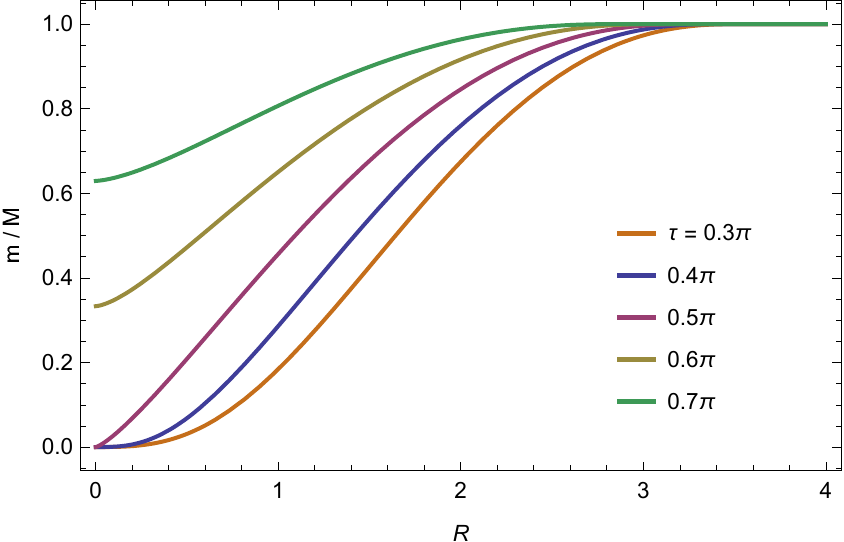} 
   \includegraphics[width=0.325\textwidth]{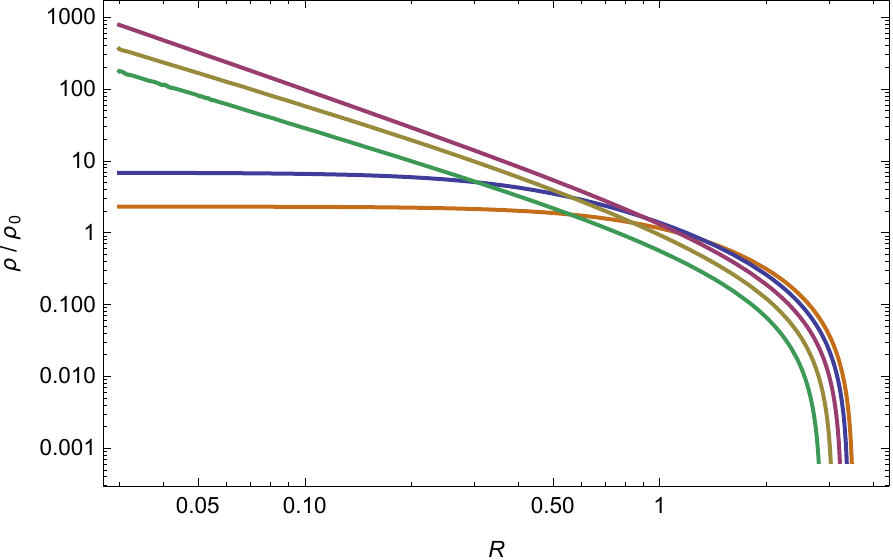} 
   \includegraphics[width=0.325\textwidth]{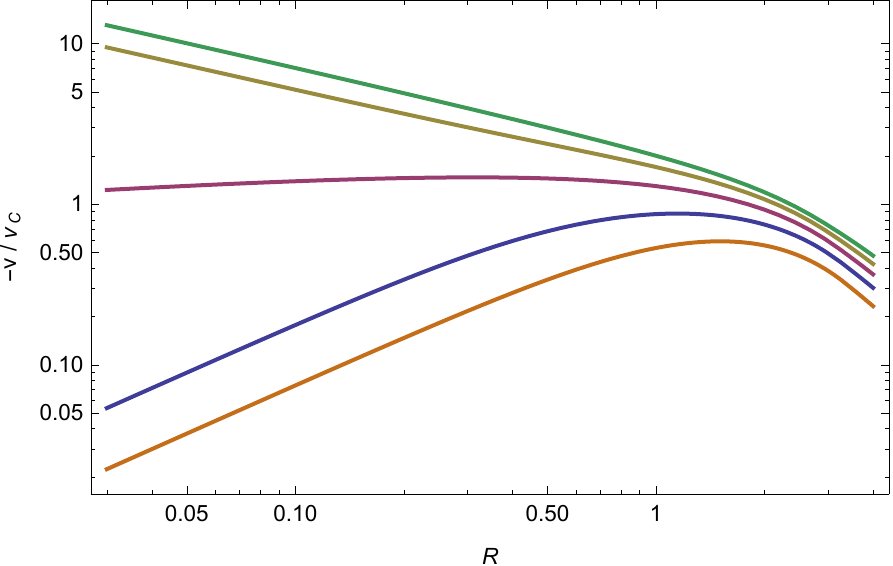} 
   \caption{The mass normalized to the total mass of the polytrope (being $M \simeq 8 M_0$) contained within $R$ (left-hand panel), the density as a function of $R$ (middle panel), and the magnitude of the infall velocity (right-hand panel) for the $\gamma = 5/3$ polytrope. At $\tau = \pi/2$, which is when the singularity forms, the density scales as $\rho \propto R^{-1.7}$ and the velocity scales as $v \propto R^{0.14}$, which are the values predicted by \citet{pen69} (see the text, specifically Section \ref{sec:conditions}, for further discussion). }
   \label{fig:gamma53}
\end{figure*}

\begin{figure*}[htbp] 
   \centering
   \includegraphics[width=0.325\textwidth]{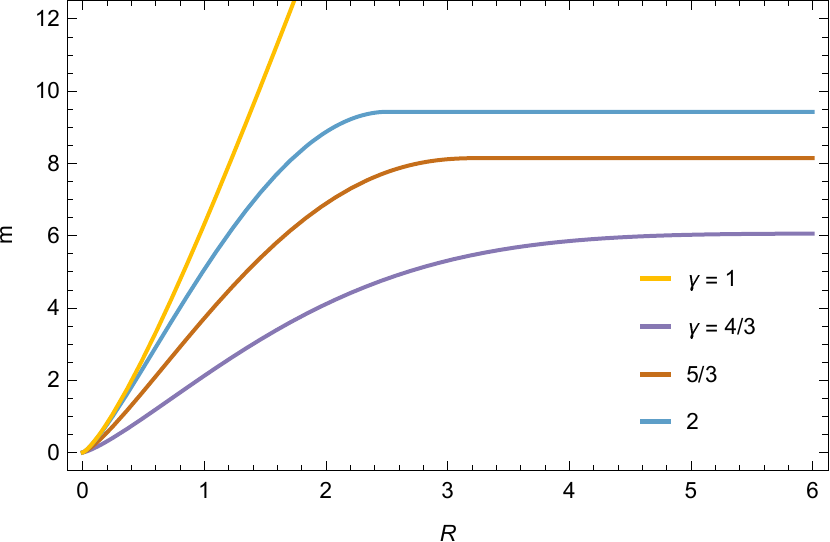} 
   \includegraphics[width=0.325\textwidth]{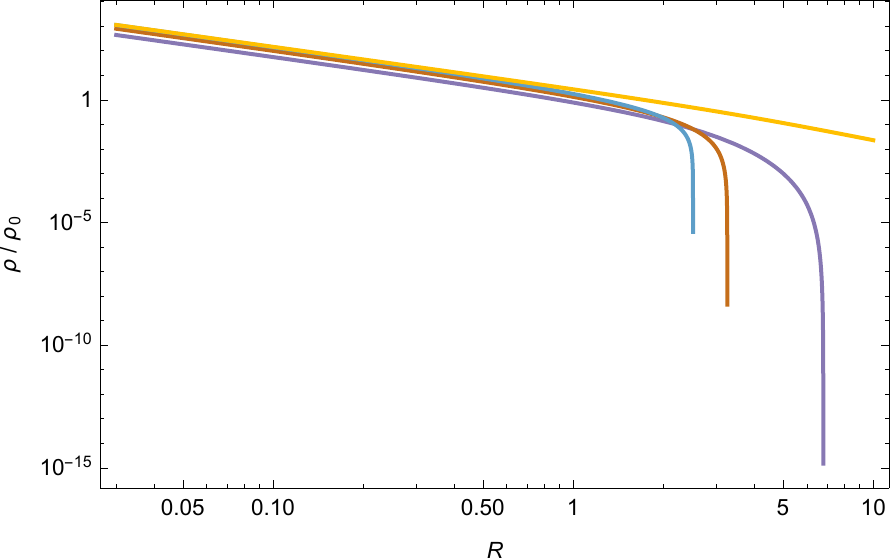} 
   \includegraphics[width=0.325\textwidth]{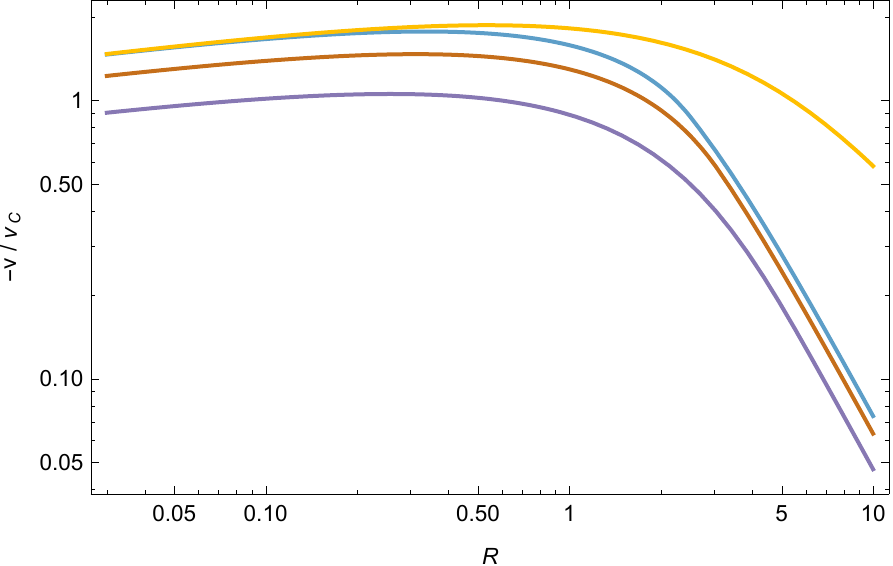} 
   \caption{The mass contained within $R$ (left-hand panel), the density (middle panel), and the magnitude of the infall velocity (right-hand panel) for the different polytropes at a time of $\tau = \pi/2$. We see that all of the polytropes form a singularity at this time, and the density and velocity scalings towards the interior are all identical. }
   \label{fig:gammascrit}
\end{figure*}

Figure \ref{fig:gamma53} shows the mass contained within $R$ (normalized to the total mass of the polytrope), the density, and the infall velocity for the case of $\gamma = 5/3$. The different curves correspond to different times since the onset of infall, with the specific times given in the legend. We see that, similar to the nearly-constant-density profiles of the previous subsection, the polytrope forms a singularity at its center at a time of $\tau = \pi/2$, and the remainder of the cloud accretes onto the point mass quite rapidly (though at a noticeably slower rate than the initial mass profiles analyzed in the previous subsection; compare the left-hand panel of Figure \ref{fig:Aplots} with the left-hand panel of Figure \ref{fig:gamma53} for times $\tau > \pi/2$). At that time, the density at small $R$ scales as $\rho \propto R^{-12/7} \simeq R^{-1.7}$, while the velocity scales as $v \propto R^{1/7} \simeq R^{0.14}$, which are not equal to the scaling relations we found for the mass profiles considered in Section 3.2. However, these power-laws are those predicted by \citet{pen69}, and we will demonstrate why this is the case in the next section. 

Figure \ref{fig:gammascrit} illustrates the normalized mass contained within $R$, the density, and the velocity for the different $\gamma$'s, all at a time of $\tau = \pi/2$. This plot demonstrates that these clouds all form a singularity at a time of $\tau = \pi/2$. In addition, while the overall normalizations are not the same, the scalings of the density and velocity for these different equations of state are all the same near the singularity. 

\begin{figure}[htbp] 
   \centering
   \includegraphics[width=0.47\textwidth]{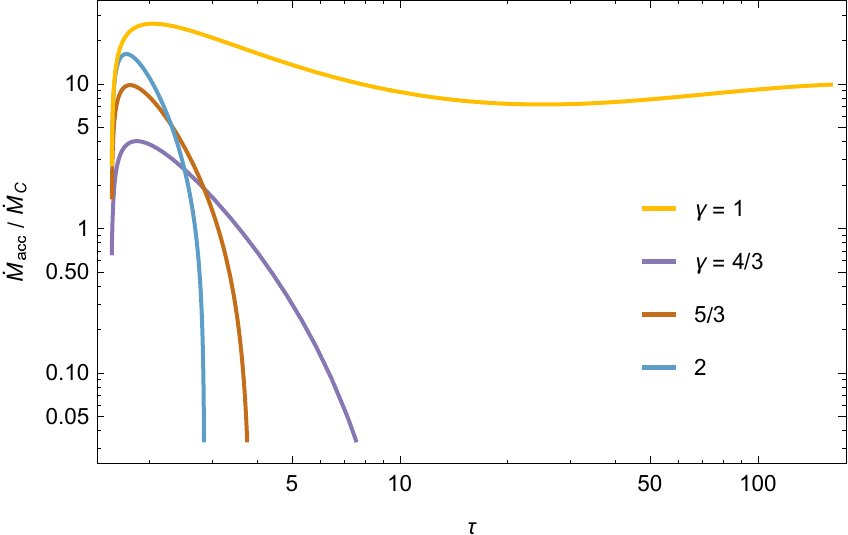} 
   \caption{The accretion rate, on a log-log scale, onto the central object as a function of time for the different polytropic indices. }
   \label{fig:mdots_poly}
\end{figure}

In Figure \ref{fig:mdots_poly} we show the accretion rates onto the central object for the different adiabatic indices. In each case we normalized the accretion rate by the total mass of the initial polytrope. In contrast to Figure \ref{fig:mdots} which showed high initial accretion rates for clouds with nearly-constant densities, the polytropic density profiles generate accretion rates that are initially zero, rise rapidly, and then decay, and the decay becomes much more pronounced as the polytropic index increases (and this result is in agreement with past findings; \citealt{ogi99}). Interestingly, the accretion rate for $\gamma = 1$ starts to decline initially, and then levels off to eventually approach a constant. Numerically we find that the isothermal accretion rate for times $\tau \gg 1$ is given by

\begin{equation}
\dot{M}_{iso} \simeq 10\dot{M}_C = 10\times\frac{2^{1/2}}{3^{3/2}}\frac{c_s^3}{G} \simeq 2.7\frac{c_s^3}{G},
\end{equation}
where $c_s^2 = kT/\mu$ is the (square of the) isothermal sound speed. Interestingly, this value is about 3 times the value of \citet{shu77}, though at earlier times the accretion rate exceeds the asymptotic value by roughly a factor of 3. 

\section{Conditions at the moment of singularity formation}
\label{sec:conditions}
The previous subsections demonstrated that, for certain initial mass profiles that corresponded to ``approximately constant'' initial densities in a cloud of gas, the center of the cloud formed a singularity in a time of $t = \pi{R_C}^{3/2}/(2\sqrt{2GM_C})$, or, in terms of time normalized by the infall time, $\tau = \pi/2$. At that instant, the density and velocity profiles of the cloud corresponded very nearly to power-laws near the cloud core, but the power-law indices depended on the initial mass profile of the cloud. The general scalings derived from our approach corresponded to Penston's $\rho \propto R^{-12/7}$ solution when the gas was described by a polytropic equation of state, but deviated from that solution otherwise. A pertinent question is then obviously: can we prove that the center of the cloud forms a singularity at $\tau = \pi/2$, and can we \emph{derive} the exact scaling laws of the density and velocity at this time?

To answer these questions, first return to the equation for the self-similar function $f$ and note that we can integrate it once to give

\begin{equation}
\xi = \frac{-\frac{i\pi}{2}-f\sqrt{f^2-1}+\ln\left(f+\sqrt{f^2-1}\right)}{(f^2-1)^{3/2}}. \label{int1}
\end{equation}
The factor of $-i\pi/2$ arises from the boundary condition $f(0) = 0$. Noting that the left-hand side of this expression tends to infinity in the limit of large-$\xi$, it follows that $f \rightarrow 1$ in this limit. By now writing $f = 1-A\,\xi^{-m}$, where $m > 0$, inserting this ansatz into Equation \eqref{int1} and keeping only lowest-order terms, we can show that

\begin{equation}
f(\xi) = 1-\frac{1}{2}\left(\frac{\pi}{2}\right)^{2/3}\xi^{-2/3}.
\end{equation}
From this solution we can obtain higher-order corrections to the function $f$ by letting $f \rightarrow f+A_n\xi^{-n}$, inserting this assumption into Equation \eqref{fss} and equating terms. Because it will be useful below, finding the next two corrections gives

\begin{equation}
f = 1-\frac{1}{2}\left(\frac{\pi}{2}\right)^{2/3}\xi^{-2/3}-\frac{1}{8}\left(\frac{\pi}{2}\right)^{4/3}\xi^{-4/3}+\frac{2^{1/3}\pi^{2/3}}{9}\xi^{-5/3}, \label{flim}
\end{equation}
with the next-highest order correction scaling as $\xi^{-2}$.

Now recall that the variable $\xi$ is given by

\begin{equation}
\xi = R^{-3/2}m^{1/2}\tau, \label{xiss}
\end{equation}
with $R$, $m$, and $\tau$ the normalized radius, mass, and time, respectively. A singularity will form at the center of the cloud at the instant when the density grows unbounded, which, recalling that $\rho \propto R^{-2}\partial{m}/\partial{R}$, will occur when $m \propto R^{\alpha}$ near the origin with $\alpha < 3$. Thus, assuming that the singularity forms at some finite time $\tau > 0$, we have

\begin{equation}
\xi \propto R^{\frac{\alpha-3}{2}} \gg 1.
\end{equation}
We emphasize that this scaling is \emph{only} valid near the origin, i.e., for $R \lesssim 1$. Using Equation \eqref{xiss} in Equation \eqref{flim}, we can show that, near the origin at the time the singularity forms, 

\begin{equation}
\frac{1-f^2}{R} = \left(\frac{\pi}{2}\right)^{2/3}m^{-1/3}\tau^{-2/3}-\frac{2^{4/3}\pi^{2/3}}{9}R^{3/2}m^{-5/6}\tau^{-5/3}, \label{1mf2}
\end{equation}
where we have kept only the first two, lowest-order terms; notice that the contribution of the term proportional to $\xi^{-4/3}$ exactly cancels here, which is why we needed the next-highest term proportional to $\xi^{-5/3}$ in our expansion. 

We can now write our fundamental equation for $m(R,\tau)$, Equation \eqref{Mofxi}, as

\begin{equation}
\frac{1-f^2}{R} = \frac{1}{M_0^{-1}(m)}, \label{feqinv}
\end{equation}
where $M_0^{-1}$ is the inverse function of the initial mass profile. Since the mass enclosed within radius $R$ becomes small as we approach the origin (even though a singularity forms, only a vanishingly small mass initially comprises that singularity), we can Taylor expand the function $M_0(m)$ about $m = 0$. Furthermore, since the density at the center of the cloud is initially a finite value, it follows that $M_0(R) \propto R^{3}$, and hence we have

\begin{equation}
M_0(m) = m^3-C\,m^{p}+\mathcal{O}(m^{q}), \label{M0eq}
\end{equation}
where $C$ is a constant, $p>3$ and $q > p$. For example, if we consider case A from Section \ref{sec:solutions}, Taylor expanding Equation \eqref{m0A} for $m_{0,A}$ gives

\begin{equation}
m_{0,A}(m) = m^3-\frac{1}{3}m^{9}+\mathcal{O}(m^{15}).
\end{equation}
Since $m \ll1$ (and hence so is $M_0$), we can show from Equation \eqref{M0eq} that the first two terms in the Taylor series of $1/M_0^{-1}(m)$ are

\begin{equation}
\frac{1}{M_0^{-1}(m)} = m^{-1/3}-\frac{C}{3}m^{(p-4)/3}.
\end{equation}
If we now use this expression for the right-hand side of Equation \eqref{feqinv} and Equation \eqref{1mf2} for the left-hand side, then we find

\begin{multline}
\left(\frac{\pi}{2}\right)^{2/3}m^{-1/3}\tau^{-2/3}-\frac{2^{4/3}\pi^{2/3}}{9}R^{3/2}m^{-5/6}\tau^{-5/3} \\ 
=m^{-1/3}-\frac{C}{3}m^{(p-4)/3}. \label{midentity}
\end{multline}
Now, this equation is valid for small, but \emph{arbitrary} values of $m$, meaning that it must be an identity in $m$. Comparing the first two terms thus yields our expression for the collapse timescale $\tau_{coll}$:

\begin{equation}
\tau_{coll} = \frac{\pi}{2}.
\end{equation}
This expression shows that, \emph{as long as the density at the center of a collapsing gas cloud is finite} (i.e., non-zero and not infinite -- two conditions we expect for any realistic collection of gas), the time at which the singularity forms is

\begin{equation}
t_{coll} = \frac{\pi}{2}\frac{R_C^{3/2}}{\sqrt{2GM_C}} = \sqrt{\frac{3\pi}{32G\rho_C}},
\end{equation}
where, as defined above, $\rho_C = 3M_C/(4\pi{R_C}^3)$ is the average density of the cloud. Thus, any clouds with the same $\rho_C$ will form a point mass at their centers, \emph{independent of any other properties of the initial distribution of the gas}.

If we now compare the next two, higher-order terms in Equation \eqref{midentity}, then we can solve for $m$ at the time the point mass forms:

\begin{equation}
m = 
\left(\frac{8}{3\pi{C}}R^{3/2}\right)^{\frac{6}{2p-3}}. \label{mapp}
\end{equation}
From the relations $\rho = 4\pi{r^2}\partial{M}/\partial{r}$ and $v = -\sqrt{2GM/r}f$, we then find that the density and velocity at $\tau = \pi/2$ are

\begin{equation}
\frac{\rho}{\rho_C} = \frac{3}{2p-3}\left(\frac{8}{3\pi{C}}\right)^{\frac{6}{2p-3}}R^{\frac{18-6p}{2p-3}}, \label{rhoapp}
\end{equation}
\begin{equation}
-\frac{v}{v_C} = \left(\frac{8}{3\pi{C}}\right)^{\frac{3}{2p-3}}R^{\frac{6-p}{2p-3}}, \label{vapp}
\end{equation}
where we recall $v_C = \sqrt{2GM_C/R_C}$. These expressions are in agreement with those found by \citet{lyn88} (see their Equation (2.4), where their $b$ defining the initial mass profile of the cloud is related to our $p$ by $b = -1+p/3$). If we now Taylor expand the initial mass configurations considered in Section \ref{sec:solutions}, then using Equations \eqref{rhoapp} and \eqref{vapp} shows that the density and velocity profiles for cases $A$, $B$, and $D$ at the time of collapse are

\begin{equation}
\frac{\rho_A}{\rho_C} = \frac{1}{5}\left(\frac{8}{\pi}\right)^{2/5}R^{-\frac{12}{5}}, \quad -\frac{v_A}{v_C} = \left(\frac{8}{\pi}\right)^{\frac{1}{5}}R^{-\frac{1}{5}},
\end{equation}
\begin{equation}
\frac{\rho_B}{\rho_C} = \frac{1}{3}\left(\frac{16}{3\pi}\right)^{\frac{2}{3}}R^{-2}, \quad -\frac{v_B}{v_C} = \left(\frac{16}{3\pi}\right)^{\frac{1}{3}}, \label{rhoBapp}
\end{equation}
\begin{equation}
\frac{\rho_D}{\rho_C} = \frac{1}{5}\left(\frac{32}{\pi^3}\right)^{\frac{2}{5}}R^{-\frac{12}{5}}, \quad -\frac{v_D}{v_C} = \left(\frac{32}{\pi^3}\right)^{\frac{1}{5}}R^{-\frac{1}{5}}.
\end{equation}

\begin{figure}[htbp] 
   \centering
   \includegraphics[width=0.47\textwidth]{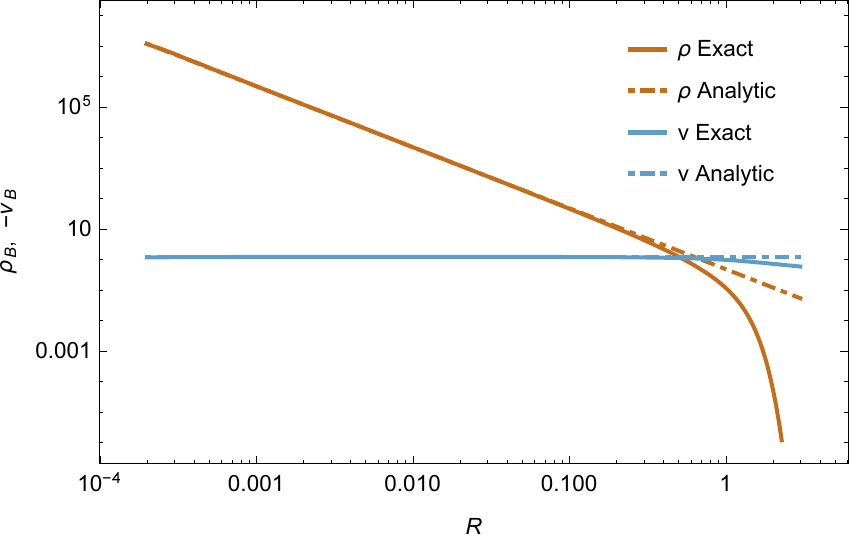} 
   \caption{The exact density (solid brown line), the approximate density (dot-dashed brown line), the exactly velocity (solid blue line), and the approximate velocity (dot-dashed blue line) at the time of collapse for case $B$ (initial mass profile of the cloud given by Equation \ref{m0B}). The density and velocity are each normalized by $\rho_C$ and $v_C$, respectively, and the exact solutions are given by Equation \eqref{rhoBapp}.}
   \label{fig:rhoBvB}
\end{figure}

Figure \ref{fig:rhoBvB} shows the normalized density (brown curves) and the normalized velocity (blue curves) on a log-log scale for the initial mass profile of case B (Equation \ref{m0B}) at a time of $\tau = \pi/2$ -- when the central singularity forms. The solid curves are the exact solution from Equation \eqref{Mofxi}, while the dashed lines are the approximate, analytical solutions given by Equation \eqref{rhoBapp}. It is apparent that the exact and approximate solutions agree very well for small $R$, with deviations of order unity occurring only when $R \gtrsim 1$. Similar, near-exact agreement is also found for cases $A$ and $D$.

For a polytropic density profile of adiabatic index $\gamma$, the function $M_0(R)$ cannot be written in closed-form. However, we can derive an approximate expression in the limit of small $R$; in particular, since $\theta \simeq 1$ for $R \ll 1$, we have from the Lane-Emden equation \eqref{LE}

\begin{equation}
\frac{1}{R^2}\frac{d}{dR}\left(R^2\frac{d\theta}{dR}\right) = -\theta^{\frac{1}{1-\gamma}} \simeq 1,
\end{equation}
which integrates to give

\begin{equation}
\theta \simeq 1-\frac{R^2}{6}.
\end{equation}
We then find for the mass enclosed within $R$:

\begin{multline}
m(R) = 3\int_0^{R}R^2\theta^{\frac{1}{\gamma-1}}dR  \\ \simeq 3\int_0^{R}R^2\left(1-\frac{1}{6(\gamma-1)}R^{2}\right)dR,
\end{multline}
or, evaluating the integral,

\begin{equation}
m(R) \simeq R^3-\frac{1}{10(\gamma-1)}R^{5}.
\end{equation}
We thus see that, for any collapsing cloud with an initially-polytropic mass profile of polytropic index $\gamma$, $C = 1/(10(\gamma-1))$ and $p = 5$, from which it follows that

\begin{equation}
\frac{\rho_{\gamma}}{\rho_C} = \frac{3}{7}\left(\frac{40(\gamma-1)}{3\pi}\right)^{\frac{6}{7}}R^{-\frac{12}{7}}, \quad -\frac{v_\gamma}{v_C} = \left(\frac{40(\gamma-1)}{3\pi}\right)^{\frac{3}{7}}R^{\frac{1}{7}}.
\end{equation}
We see that these scalings give, as we predicted from the numerical solutions in Section \ref{sec:solutions}, $\rho \propto R^{-12/7}$ and $v \propto R^{1/7}$. This finding shows that Penston's self-similar scaling is recovered only if $p = 5$. (For an isothermal sphere, it is easy to show from the modified Lane-Emden equation, Equation \ref{isoth}, that the factor of $\gamma-1$ is simply replaced by $1$). 

From a physical standpoint we expect the gravitational field towards the center of a collapsing cloud of gas to be zero. Since the gravitational field is just $g = -GM/r$, it follows that $p > 4$. Using this value of $p$ in Equations \eqref{rhoapp} and \eqref{vapp}, we see that the density at the time a singularity forms must fall off faster (i.e., have a more negative exponent) than

\begin{equation}
\rho_L \propto R^{-\frac{6}{5}},
\end{equation}
while the radial velocity must fall off faster than

\begin{equation}
v_L \propto R^{\frac{2}{5}}.
\end{equation}

\subsection{Accretion rates}
Using the continuity equation to write $\dot{m} = -v\,\partial{m}/\partial{R}$, Equations \eqref{mapp} and \eqref{vapp} show that the accretion rate onto the point mass at the time it forms is

\begin{equation}
\dot{m} \propto R^{\frac{3(6-p)}{2p-3}}. \label{mdotofp}
\end{equation}
This expression illustrates an interesting dichotomy: if $p < 6$, the accretion rate onto the point mass -- obtained by letting $R \rightarrow 0$ -- goes to zero at the time the point mass forms; if $p > 6$ the initial accretion rate grows unbounded; \emph{and it is only if $p \equiv 6$} that the accretion rate remains finite at the time of singularity formation. Investigating Figures \ref{fig:mdots} and \ref{fig:mdots_poly}, we see that this finding is confirmed, as cases A and D ($p = 9$) have asymptotically growing initial mass accretion rates, case B ($p=6$) retains a finite accretion rate, and the accretion rates of the polytropes ($p = 5$) go to zero. 

Figure \ref{fig:mdots_poly} also shows that, for an isothermal sphere, the accretion rate of the central object appears to approach a constant for $\tau \gg 1$. To demonstrate that the asymptotic solution is in fact a constant accretion rate, we note that, for $R \gg 1$, the initial density profile of the cloud satisfies $\rho_0(R) \simeq 2/R^2$ (the factor of 2 can be found by inserting $\theta = Ar^{-2}$ in Equation \ref{isoth} and solving for the value of $A$), and hence $M_0(R) \simeq 6R$. For times $\tau \gg 1$, it follows that $\xi = m^{1/2}R^{-3/2}\tau \gg 1$, and hence, from Equation \eqref{flim}, $1-f^2 \simeq (\pi/2)^{2/3}\xi^{-2/3}$. Using this result in Equation \eqref{Mofxi} then gives

\begin{equation}
m \simeq 6\left(\frac{\pi}{2}\right)^{-2/3}R\left(m^{1/2}R^{-3/2}\tau\right)^{2/3},
\end{equation}
and making some simple rearrangements yields

\begin{equation}
m \simeq \frac{2\times6^{3/2}}{\pi}\,\tau \simeq 9.4\,\tau,
\end{equation}
in good agreement with the estimate given at the end of Section \ref{sec:solutions} for a $\gamma = 1$ polytrope (the oscillation in the accretion rate shown in Figure \ref{fig:mdots_poly} is due to the fact that the density of an isothermal sphere does not fall off exactly as $\rho \propto r^{-2}$, but periodically over and under-estimates this scaling). We can further generalize this result by letting the initial mass profile scale as $M_0 = b\,R^{\,n}$ in the limit of $R \gg 1$, where $b$ and $n$ are constants. In this case, following the same steps that we did for $n = 1$ shows that

\begin{equation}
m \simeq \left[b\left(\frac{\pi}{2}\right)^{-\frac{2n}{3}}\right]^{\frac{3}{3-n}}\tau^{\frac{2n}{3-n}}. \label{mdotofn}
\end{equation}
Interestingly, this expression is always independent of $R$, no matter the value of $n$ (although, strictly speaking, this relation is only valid in the limit that $\xi \gg 1$, and very large values of $R$ will start to violate this assumption). Also, since $f \simeq 1$ in this limit, we have $v = -\sqrt{2Gm/R}$, and hence rearranging the continuity equation gives

\begin{equation}
\rho \propto R^{-3/2}\tau^{\frac{2n-3}{3-n}}. \label{rhoofn}
\end{equation}
This expression shows that $n = 3/2$ is the unique value for an infinite-mass envelope that approaches time-independence in its density profile, and thus represents the limit where the mass fed into the accreting region is balanced by the accretion onto the point mass. For $n < 3/2$, the mass supply rate is shallower than the accretion rate, causing the density to be a decreasing function of time. Finally, if $n > 3/2$, so much mass is provided to the protostar per unit time that its accretion rate cannot compensate, resulting in an increasing density with time.

Much effort has been dedicated to the study of clouds with an isothermal equation of state, which has $\rho \propto r^{-2}$ and, as we demonstrated above, a constant mass accretion rate onto the central point mass . However, authors have also proposed the ``logotropic'' pressure profile, which satisfies $\rho \propto r^{-1}$ for $r \gg 1$ \citep{mcl96}, that includes the effects of magnetic turbulence on the pressure and, correspondingly, density profile of the initial collapsing cloud (this has also been seen in some observations; e.g., \citealt{cer85, van00}). For this logotropic density profile, $m \propto r^{2}$, and hence the mass of the central object grows as $m \propto \tau^4$ -- much steeper than the constant rate predicted for $\rho \propto r^{-2}$ (this is also in agreement with \citealt{mcl97}). In this case, the density near the accreting protostar is an \emph{increasing} function of time, satisfying $\rho \propto R^{-3/2}\tau$.

\begin{figure*}[htbp] 
   \centering
   \includegraphics[width=0.495\textwidth]{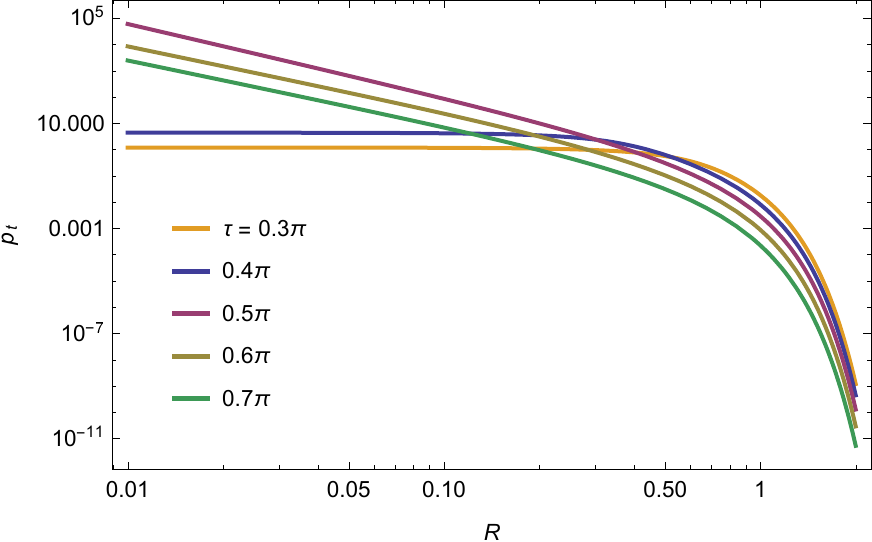} 
   \includegraphics[width=0.495\textwidth]{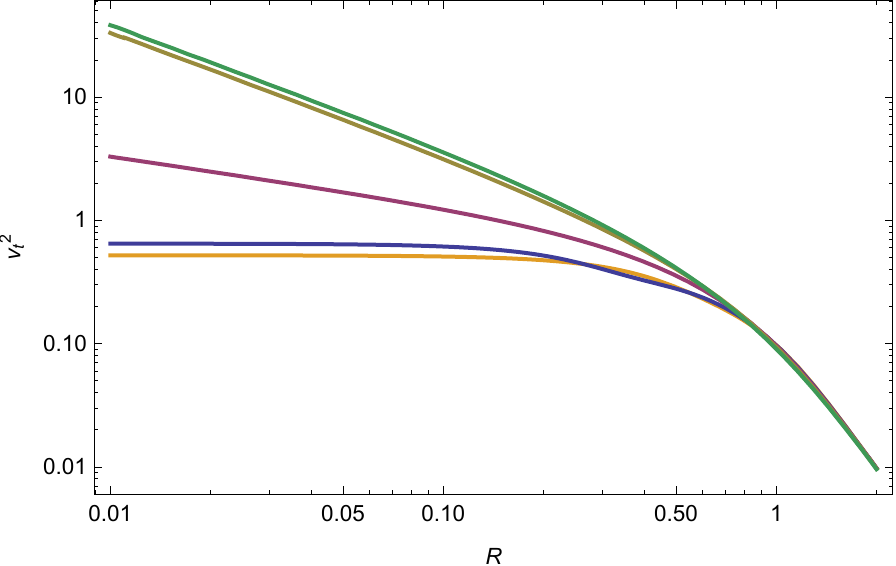} 
   \caption{The turbulent pressure (left panel), found by integrating Equation \eqref{press} from infinity to $R$ (assuming that the turbulent pressure vanishes at infinity), and the square of the turbulent velocity, $v_t^2 \equiv p/\rho$, for the initial density profile appropriate to case A in Section \ref{sec:solutions} (see Equation \ref{m0A}).}
   \label{fig:turbs}
\end{figure*}

In reality, we of course do not expect the molecular gas surrounding any individual star-forming core to exactly follow a single power-law. Figure \ref{fig:mdots_poly} shows that this lack of a single power-law is in fact reflected in the accretion rate: the initial decline in $\dot{M}_{acc}$ for $\gamma = 1$ occurs because the density profile of a polytropic sphere temporarily falls off steeper $\rho \propto r^{-2}$, and the accretion rate increases slightly as the density profile declines in a manner that is slightly shallower than $\propto r^{-2}$. 

\section{Turbulent pressure}
\label{sec:pressure}
So far we have ignored the contribution of the pressure of the gas to resisting the infall of the material. Retaining a finite pressure $p$, the radial momentum equation takes the form

\begin{equation}
\frac{\partial{v}}{\partial{t}}+v\frac{\partial{v}}{\partial{r}}+\frac{1}{\rho}\frac{\partial{p}}{\partial{r}} = -\frac{GM}{r^2}.
\end{equation}
From this expression, we see that the sound speed (squared) $c_s^2 \sim p/\rho$ contributes an additional velocity that enters into the evolution of the collapsing cloud. For the case of a general pressure gradient, then, it is unlikely that there remain exact solutions for arbitrary initial densities of the infalling gas.

Nevertheless, there is one specific type of pressure support that permits general solutions of the kind found in the previous sections: if we let

\begin{equation}
\frac{1}{\rho}\frac{\partial{p}}{\partial{r}} = -\left(1-\beta^2\right)\frac{GM}{r^2}, \label{press}
\end{equation}
where $\beta \lesssim 1$ is a constant, then the effective velocity and timescale are simply replaced by $v \rightarrow\beta{v}$ and $\tau \rightarrow \tau/\beta$. Therefore, the solutions in this case correspond to precisely the same solutions found for pressureless collapse, but with a longer timescale and a slower infall velocity. 

Equation \eqref{press} corresponds to the scenario in which the speed associated with pressure perturbations scales as the local infall speed, i.e., $p/\rho \sim c_s^2 \sim GM/r$. While the thermal pressure of the gas may not conform automatically to this prescription for any arbitrary gas cloud, especially during the initial isothermal phase of collapse, this situation may be actualized when turbulent gas motions provide the feedback that resists the gravitational field. In this case, the sound speed should be thought of as the RMS value of the turbulent speed. It was recognized early on that, because of the large observed line widths, turbulence may play a large role in dictating the evolution of star-forming regions \citep{ful92}. Furthermore, if the turbulence in the cloud is driven by the conversion of gravitational energy into turbulent motions \citep{rob12}, then one naturally expects the pressure to scale as Equation \eqref{press} \citep{mur15}.

Figure \ref{fig:turbs} shows the turbulent pressure, found by integrating Equation \eqref{press} from infinity to $R$ and assuming $p(R\rightarrow\infty) = 0$, in the left-hand panel, and the square of the turbulent speed, $v_t^2 = p/\rho$, in the right-hand panel for the initial mass profile $m_{0,A}$ given by Equation \eqref{m0A}. The different curves correspond to the different times as indicated in the legend. We see that the pressure and velocity both approach power-laws at the moment that the point mass forms, and we can determine those power-laws by using the scalings for the mass and density found in the previous section, namely Equations \eqref{mapp} -- \eqref{vapp}, in Equation \eqref{press}. Doing so gives

\begin{equation}
p \propto R^{\frac{30-8p}{2p-3}},
\end{equation}
and using $p = 9$ -- the value appropriate to case $A$ -- gives $p \propto R^{-14/5}$, which is slightly steeper than the scaling $p \propto R^{-2.5}$ that we expect for times after the singularity forms (this comes from using $m \sim 1$ and $\rho \sim r^{-3/2}$ in Equation \ref{press}); this is apparent by-eye from Figure \ref{fig:turbs}. Using the same scalings, we find that the square of the turbulent velocity scales as

\begin{equation}
v_t^2 \propto R^{\frac{12-2p}{2p-3}},
\end{equation} 
which gives $v_t^2 \propto R^{-2/5}$ for $p = 9$ -- in exact agreement with Figure \ref{fig:turbs}. For late times, we recover that the pressure scales as $R^{-5/2}$ and the turbulence velocity scales as $R^{-1/2}$, showing that the gravitational field of the point mass has dominated the dynamics of the inflowing gas.

It is possible to generalize our prescription for the turbulent pressure given in Equation \eqref{press} by letting $\beta \rightarrow \beta(\xi)$. In this case, making the same assumption about the scaling of the velocity, i.e., $v = \sqrt{2GM/r}f(\xi)$ with $\xi = \sqrt{2GM}t/r^{3/2}$, results in a slightly altered equation for the function $f$: 

\begin{equation}
f' = \frac{\beta^2-f^2}{2+3f\xi}.
\end{equation}
Similarly, the solution for $M(t,r)$ is given by

\begin{equation}
M = M_0\left(r\exp\left\{\int_0^{\xi}\frac{2f'(y)}{\beta(y)^2-f(y)^2}dy\right\}\right), \label{masseq}
\end{equation}
where $y$ is a dummy variable of integration.

In principle there is no restriction on the functional dependence of $\beta(\xi)$. However, if one interprets the pressure here as arising from turbulent motions of the gas, then these motions should be self-consistently generated from the gravitational infall of the cloud. Therefore, one might expect a relation similar to Equation (15) of \citet{mur15} as the additional ``energy'' relation -- responsible for converting gravitational energy into turbulent motions -- that closes the system and establishes the specific form of $\beta$. In this case, the assumption of the self-similarity of $\beta$ and $f$ would generate two coupled, ordinary differential equations that one could solve numerically, leaving Equation \eqref{masseq} to determine the mass as a function of time. We leave the further analysis of this possibility to a future investigation.

\section{Overdensities}
\label{sec:overdensities}
In Section \ref{sec:solutions} we considered the infall of clouds with (barring the test case of a constant-density sphere) smooth, monotonically-decreasing density profiles. However, these highly-idealized scenarios are unlikely to be actualized in realistic astrophysical environments, especially with the emerging paradigm of the turbulent nature of star formation. In contrast to the canonical, slowly-progressing infall of an isothermal sphere (e.g., \citealt{shu77}), observations and simulations (e.g., \citealt{fed15, fed15b, mur15b}) suggest that the initial formation scenario is much more violent and dynamical, with the collisions of filaments marking the cites of the growth of stellar cores.

In light of this viewpoint, it is likely the case that there will be regions of anomalously high density within the collapsing cloud, potentially generated by the passage of shocks as material compresses. The evolution of such overdensities as they collapse inward with the rest of the core will then proceed in a non-trivial manner, as can be seen from the fact that, if the mass of the overdensity is sizeable in comparison to the mass contained within the rest of the cloud and its extent is relatively small, the gravitational force on the outer edge of the overdensity will be larger than that on the inner edge. The outer edge will therefore accelerate towards the inner edge, increasing the magnitude of the overdensity relative to the rest of the collapsing region. On the other hand, if the mass contained interior to the overdensity is large in comparison to the mass of the overdensity itself, then tides will tend to shear out the overdensity.

\begin{figure*}[htbp] 
   \centering
   \includegraphics[width=0.325\textwidth]{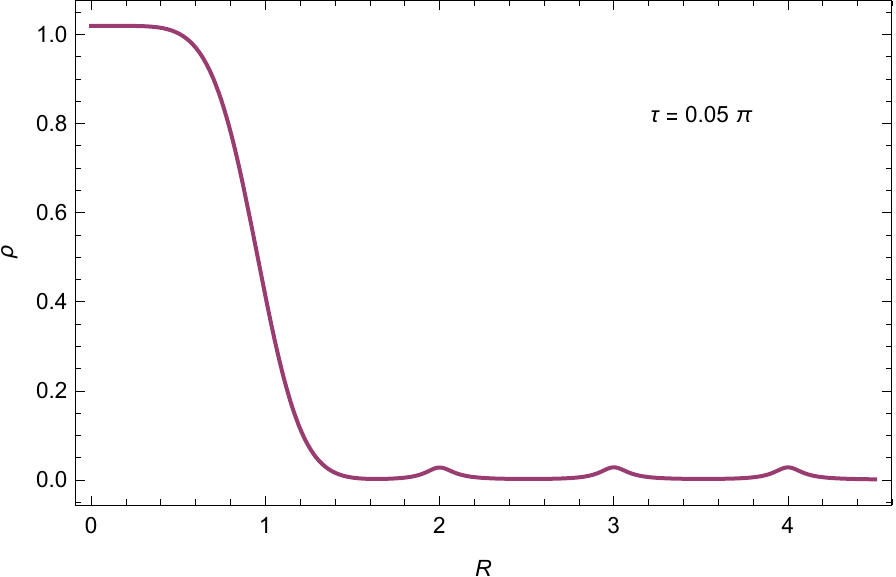} 
   \includegraphics[width=0.325\textwidth]{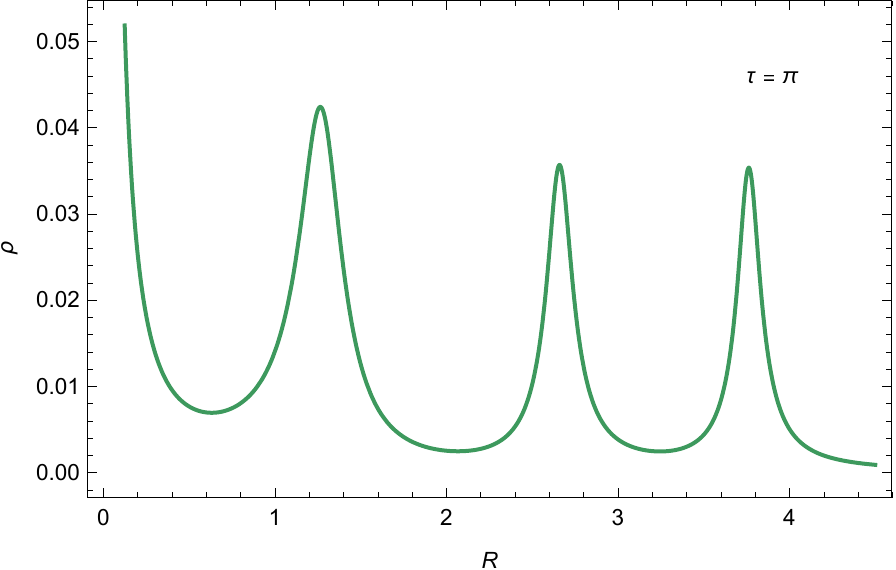} 
   \includegraphics[width=0.325\textwidth]{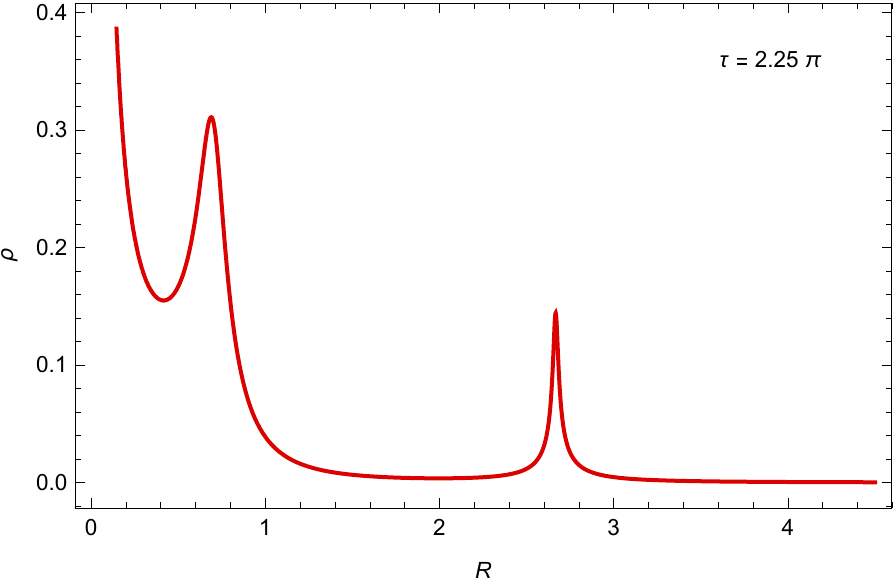} 
      \includegraphics[width=0.325\textwidth]{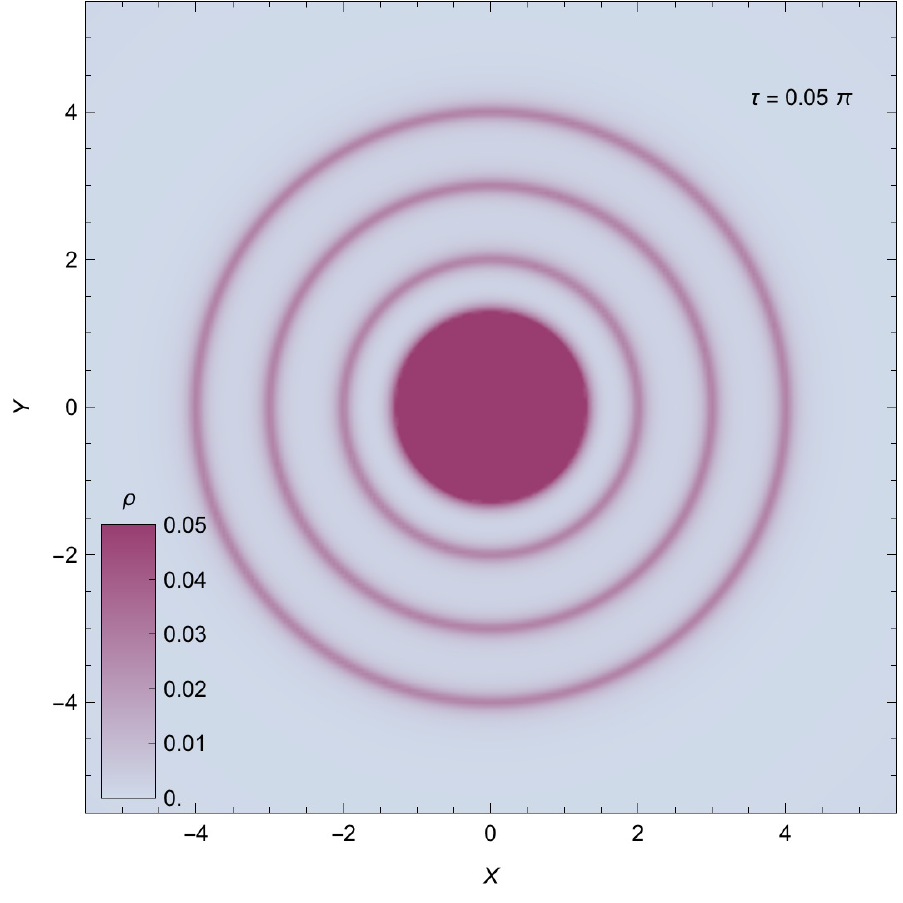}
   \includegraphics[width=0.325\textwidth]{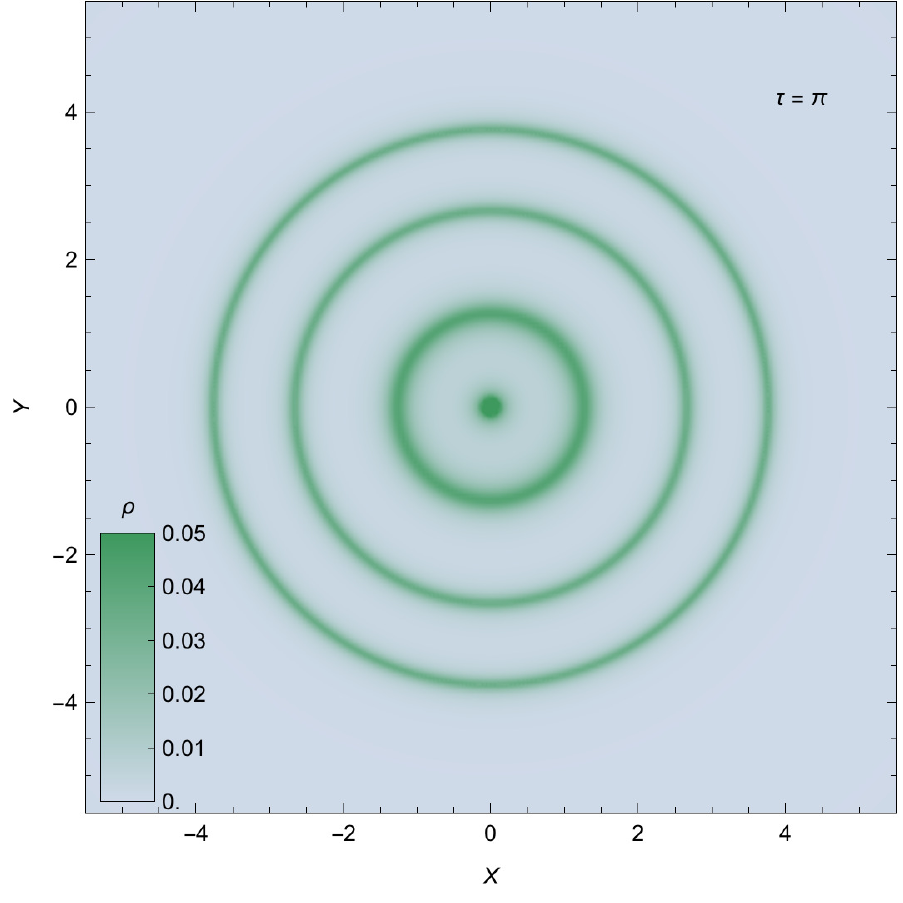}
   \includegraphics[width=0.325\textwidth]{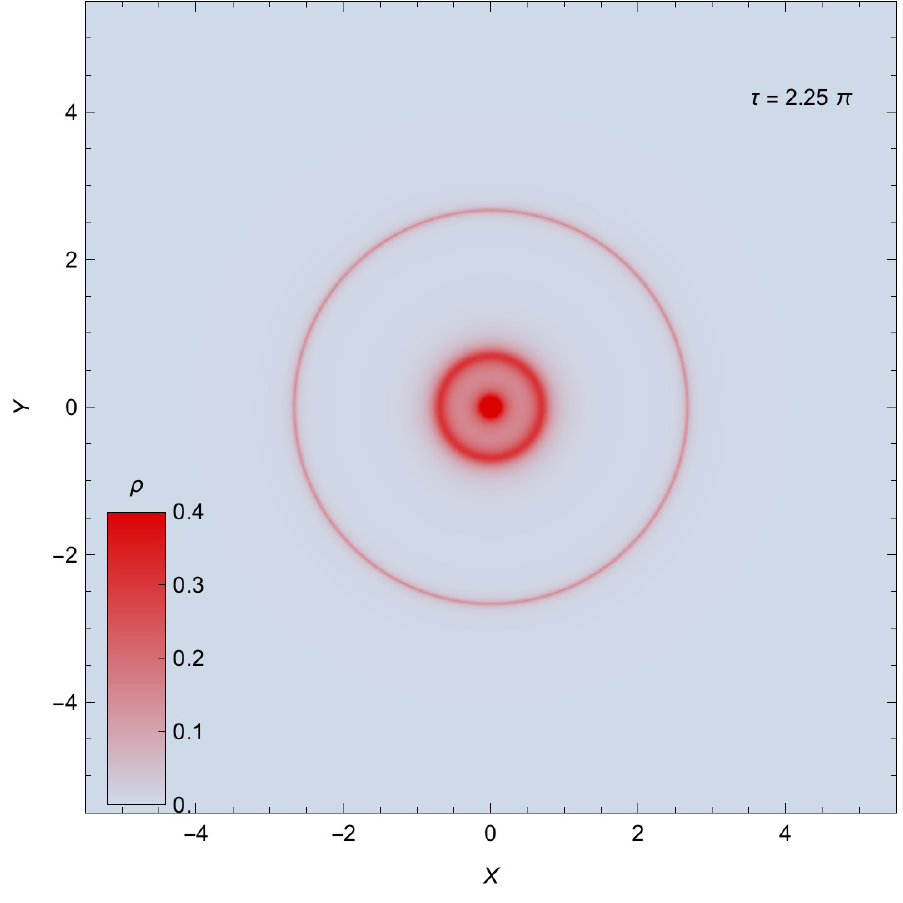}
   \includegraphics[width=0.325\textwidth]{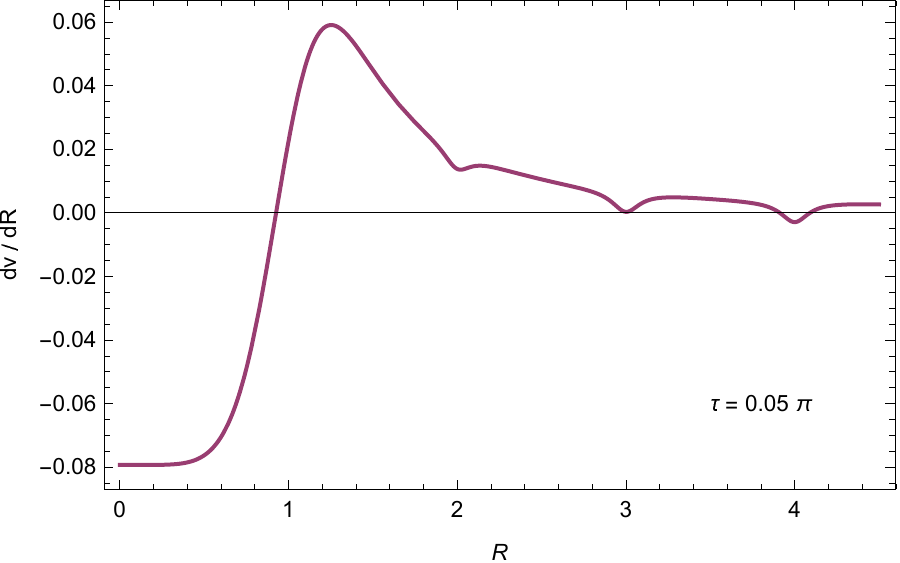} 
   \includegraphics[width=0.325\textwidth]{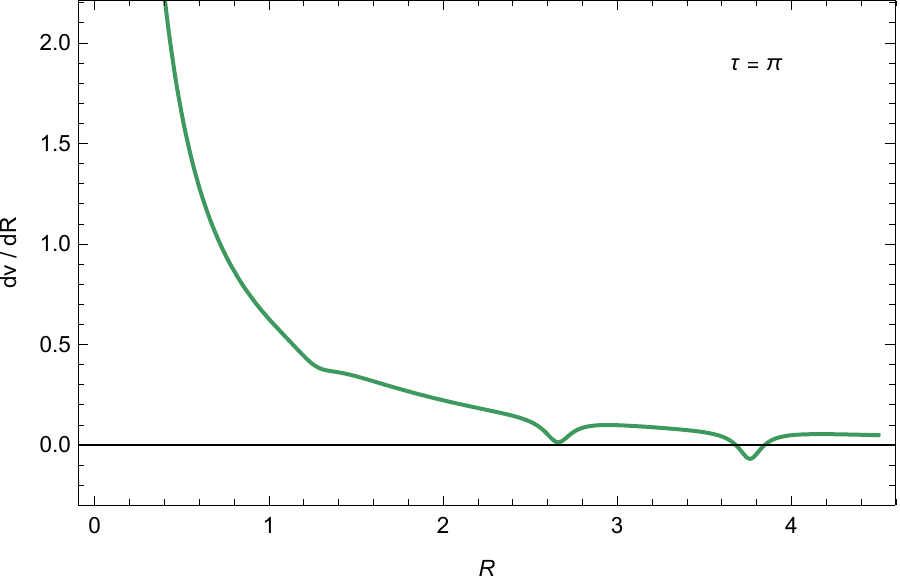}
   \includegraphics[width=0.325\textwidth]{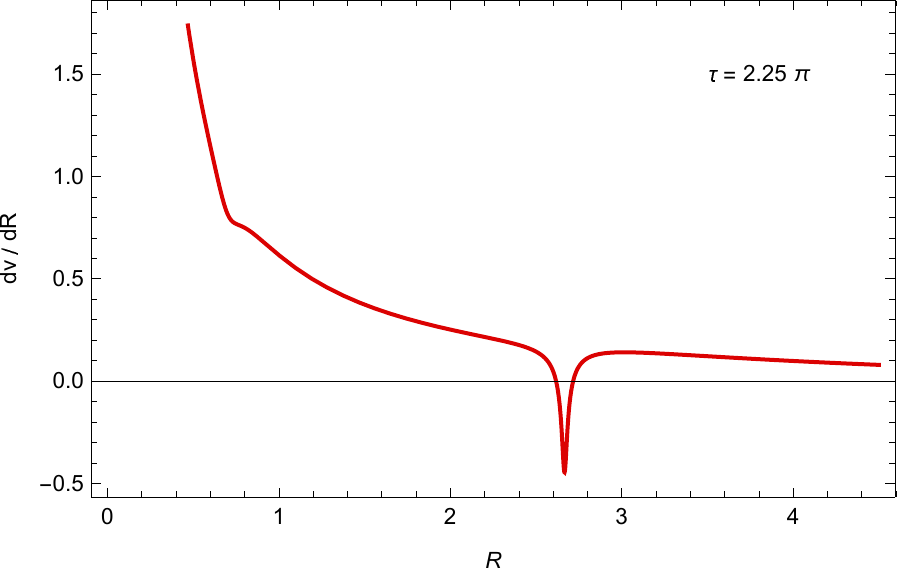}
   \caption{Top row: the density profile of a collapsing cloud, including three perturbations at initial radii of $r_0 = 2$, 3, and 4, at a time of $\tau = 0.05\pi$ (just after the start of the infall), $\tau = \pi$, and $\tau = 2.25\pi$. Middle row: a contour plot of the density, with the color bar indicating the correspondence between the color and the value of the density, for the same times as the top row, which gives a direct visual representation of the growth or depletion of the each overdensity. Bottom row: the velocity gradient, $dv/dR$, normalized by $R_C/v_C$, for the same times. As discussed on the text, regions of the flow satisfying $dv/dR < 0$ are collapsing under their own self-gravity, which gives a measure of the gravitational stability or instability of the overdensities.}
   \label{fig:overdensities}
\end{figure*}

We can characterize the criterion that differentiates between these two scenarios by writing the gravitational forces on the inner and outer edges of the overdensity as

\begin{equation}
f_{in} = \frac{GM_{in}}{r_{in}^2},
\end{equation}
\begin{equation}
f_{out} = \frac{GM_{out}}{r_{out}^2} \simeq\frac{GM_{in}}{r_{in}^2}\left(1+\frac{\delta{M}}{M_{in}}-2\frac{\delta{r}}{r_{in}}\right), \label{fout}
\end{equation}
where the final line follows from the assumption that the width of the overdensity, $\delta{r}$, is small compared to $r_{in}$, and that the mass contained in the overdensity, $\delta{M}$, is small compared to $M_{in}$. Since runaway collapse will occur if $f_{out} > f_{in}$, we see that this condition is satisfied if

\begin{equation}
\frac{\delta{M}}{M_{in}} > 2\frac{\delta{r}}{r_{in}}. \label{mcriter}
\end{equation}
Writing $\delta{M} \simeq 4\pi{}r_{in}^2\rho_1\delta{r}$ and $M_{in} = 4\pi{}r_{in}^3\langle{\rho}\rangle/3$, where $\rho_1$ is the average density within the overdense region and $\langle\rho\rangle$ is the average density within $r_{in}$, this condition becomes

\begin{equation}
\rho_o \gtrsim \frac{2}{3}\bar\rho. \label{rhocriter}
\end{equation}
If this condition is satisfied, the outer edge and inner edges of the overdensity will converge, leading to its runaway growth.

We will assess the validity of this criterion by letting the initial density profile of the collapsing cloud be given by

\begin{equation}
\rho_0(r) = \bar\rho(r)+\rho_1\frac{\frac{R}{R_0}}{1+\left(\frac{r-r_0}{\sigma}\right)^2}, \label{rhotot}
\end{equation}
where $\bar\rho$ is the density profile of an unperturbed cloud (such as those considered in Section \ref{sec:solutions}), and $\rho_1$, $r_0$, and $\sigma_r$ are the respective magnitude, location, and half-width of the overdensity; the factor of $r/r_0$ simply ensures that the magnitude of the overdensity equals zero at the origin.

The top-left Figure \ref{fig:overdensities} shows an initial density profile with $\bar\rho(r)$ equal to that of case A analyzed in Section \ref{sec:solutions}, i.e., $\bar\rho(R,t=0) = \text{sech}^2(R^3)$ (technically-speaking, this is at a time of $0.05\,\pi$, or just after the cloud has started to collapse). On top of this we introduced three perturbations of equal magnitude $\rho_1 \simeq 0.03$ and equal width $\sigma \simeq .1$, with $r_0 = 2$, 3, and 4. With this density profile, the mass contained within the overdensities at positions $r_0 = 2$, 3 and 4 are $\delta{M}_2 \simeq 0.05$, $\delta{M}_3 \simeq 0.12$, and $\delta{M}_4 \simeq 0.22$ (note that, even though the magnitude of the density is the same, the larger value of $r_0$ implies a greater mass contained within the overdensity due to the factor of $r_0^2$ when converting from the density to the mass), and thus with $\delta{r} = 2\sigma \simeq 0.2$, we have $\delta{M}_2/\delta{r}\simeq 0.25$, $\delta{M}_3/\delta{r} \simeq 0.61$, and $\delta{M}_4/\delta{r} \simeq 1.1$. Using $M_{in} \simeq 1$, $r_{in} = r_0$ appropriate to each respective overdensity, we find $\delta{M}_2/\delta{r} < 2M_{in}/r_{in}$, $\delta{M}_3/\delta{r} \simeq 2M_{in}/r_{in}$, and $\delta{M}_4/\delta{r} > 2M_{in}/r_{in}$. Thus, according to the self-gravitating criterion given by Equation \eqref{mcriter}, we see that $\delta{M}_2$ should be shear dominated, $\delta{M}_3$ should be just on the cusp of self-gravitating, and $\delta{M}_4$ should collapse under its own self-gravity.

The middle and right-hand panels of the top row of Figure \ref{fig:overdensities} show the density at times of $\tau = \pi$ and $\tau = 2.25\pi$. In the middle panel we see that the innermost overdensity is clearly torn apart by the tidal shear, and in the right panel it is apparent that the second perturbation is starting to be tidally-destroyed while the outermost perturbation is collapsing under its own self-gravity into an infinitely-thin shell. The middle row shows the contour plots of the density at these same times, with the colors indicating the magnitude of the density, as indicated in the legend (values of the density that exceed the maximum value in the legend are indicated by the darkest color). This representation more directly quantifies the degree to which the overdensities sharpen or dissipate, as the width of the isocontour is directly related to the width of the perturbation.

The bottom row of Figure \ref{fig:overdensities} shows the derivative of the radial velocity, $dv/dR$, throughout the collapsing region. This quantity gives us an instantaneous measure of the degree to which the perturbations are collapsing under their own self-gravity: the width of the perturbation will decrease if the velocity of its outer edge is more negative than the velocity of its inner edge. Denoting these respective velocities by $v(r+\delta{r})$ and $v(r)$, then the condition for runaway growth of the overdensity is $v(r+\delta{r}) - v(r) \simeq dv/dr\delta{r} < 0$. We see that these panels directly confirm our predictions about the self-gravitating nature of the density perturbations: the velocity gradient within the first overdensity is always positive, showing that the tidal shear is preventing the perturbation from growing. Only the densest part of the second overdensity is close to having a negative velocity gradient, showing that it is just on the edge of self-gravitating. Finally, the outermost perturbation satisfies $dv/dR < 0$ over most of its extent, showing that it will collapse under its own self-gravity.

\begin{figure}[htbp] 
   \centering
   \includegraphics[width=0.47\textwidth]{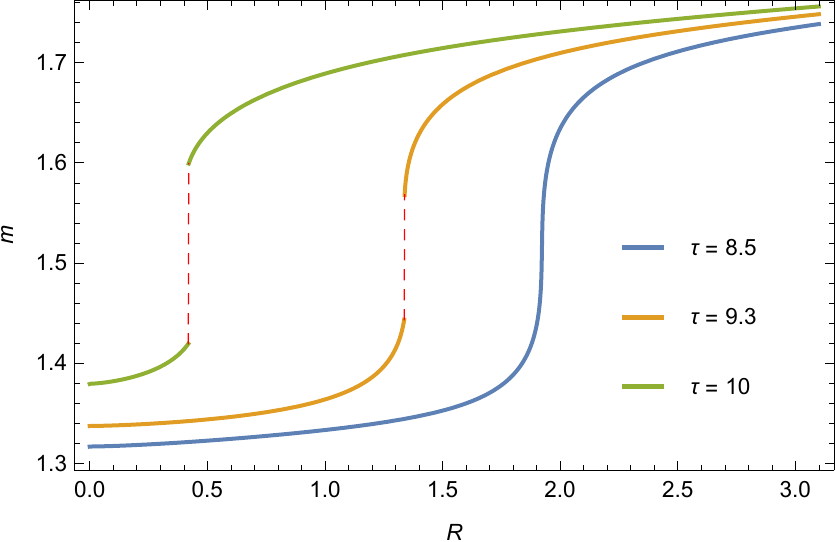} 
   \caption{The mass contained within $R$ at three different times, shown in the legend, when the overdensity originally at $r_0 = 4$ starts to collapse under its own self-gravity. Vertical, dashed red lines show where the discontinuity occurs, demonstrating that the overdensity has formally collapsed into an infinitely-thin shell.}
   \label{fig:mcaustic}
\end{figure}

The overdensity satisfying the self-gravitating criterion continues to collapse in on itself until its density approaches a caustic, becoming an infinitely-thin shell at later times. We can estimate the time it takes for this to occur by noting that, if the shell is self-gravitating, then the force that causes it to collapse on itself is, from Equation \eqref{fout}, $\Delta{f} \simeq G\delta{M}/r_{in}^2$. Setting this force equal to $d^2\delta{r}/dt^2$ then gives

\begin{equation}
\tau_{sh} \simeq 2R_{in}\left(\frac{\delta{r}}{\delta{M}}\right)^{1/2}, \label{taush}
\end{equation}
and this time is now normalized to $T_C = R_C^{3/2}/\sqrt{2GM_C}$. For the outermost perturbation, this timescale corresponds to approximately $\tau_{sh} \simeq 8$.  

Figure \ref{fig:mcaustic} shows the mass contained within radius $R$ at the three different times indicated in the legend, which are after the first two perturbations have been accreted by the point mass. The third perturbation, which was originally at $r_0 = 4$, has moved in significantly closer to the origin by this time, and it is apparent from the sharp rise in $m(R)$ at a time of $\tau = 8.5$ -- slightly greater than the time predicted in Equation \eqref{taush} -- that its density profile is becoming increasingly concentrated. For times shortly thereafter, the central portion of the overdensity formally collapses to a singularity, with $m(R)$ discontinuously traversing the newly-formed shell (shown by the vertical dashed lines). 

\begin{figure}[htbp] 
   \centering
   \includegraphics[width=0.47\textwidth]{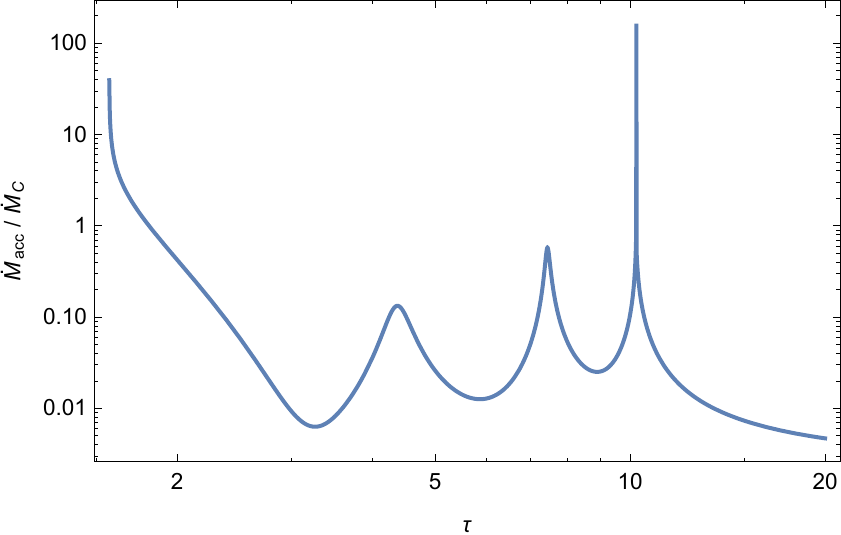} 
   \caption{The accretion rate onto the central point mass as a function of time; the spikes at discrete times coincide when the initial density perturbations fall to the origin.}
   \label{fig:mdotcaustic}
\end{figure}

Figure \ref{fig:mdotcaustic} illustrates the accretion rate onto the point mass at the origin. The very steep dropoff at $\tau = \pi/2$ is expected, as the very flat initial density profile appropriate to case $A$ has $p = 9$ (cf.~Equation \ref{mdotofp}; also see Figure \ref{fig:mdots}). The peaks at later times correspond to the accretion of the density perturbations. At times greater than roughly $\tau \simeq 10$, the accretion rate appears to drop off roughly as a power-law; however, at very late times ($\tau \gtrsim 100$, not shown in the figure), the rate upturns again and follows approximately $\dot{m} \propto \tau^3$. The reason for this behavior is that the density perturbations fall off as, from Equation \eqref{rhotot}, $\rho_1 \propto 1/R$ for $R \gg 1$, and hence $m(r) \propto R^2$ in this limit. Thus, according to Equation \eqref{mdotofn}, the mass contained within $R$ at late times should scale $m \propto \tau^4$, confirming the accretion rate of $\dot{m} \propto \tau^3$. 

The formation of an infinitely-thin shell may be an artifact of the neglect of gas pressure within our calculations, as the converging flow will eventually increase the thermal energy to the point where it will start to become relevant in resisting the velocity gradient. Nevertheless, if the  pressure gradient is originally unimportant and the flow is still in the phase where it can cool efficiently, then the condition for self-gravity to overwhelm the tidal shear -- permitting the onset of runaway growth of any preexisting overdensities within the collapsing region -- and the timescale over which the overdensities grow into shells, given by Equation \eqref{taush}, are likely accurate for realistic star-forming or protogalactic clouds. 

\section{Summary and Conclusions}
\label{sec:conclusions}
In this paper we considered the infall of a pressureless, spherically-symmetric gas cloud under its own self-gravity. Contrary to past investigations, we did not make any assumption about the self-similar state of the infall, and merely made a convenient change of variables to deduce the general solution to the problem that is valid for any initial density profile of the cloud. Specifically, writing the infall velocity as $v_r = -\sqrt{2GM/r}f(\xi)$, where $\xi = \sqrt{2GM}t/r^{3/2}$, and using this form in the radial momentum equation \eqref{rmom1} results in an exact, ordinary differential equation for the function $f(\xi)$ (Equation \ref{fss}), and Equation \eqref{Mofxi} determines the mass enclosed within any given radius at time $t$ with an initial mass profile $M_0(R)$ (an initial density profile $\rho_0 = 3/(4\pi{R}^2)\,dM_0/dR$). We also demonstrated, via a proof by contradiction (Section \ref{sec:proof}), that these are the \emph{unique} solutions to the problem, despite the nonlinear nature of the equations.

Since the variable $\xi$ is defined in terms of the mass $M(r,t)$, Equation \eqref{Mofxi} is an implicit relation for $m(r,t)$. However, since the function $f(\xi)$ can be determined numerically to a high degree of accuracy from Equation \eqref{fss}, Equation \eqref{Mofxi} can simply be regarded as an algebraic one. We have thus transformed the nonlinear, coupled, partial differential equations given by Equations \eqref{rmom1} and \eqref{cont2} into an ODE (that can actually be integrated once; see Equation \ref{int1}) and an algebraic equation.

We provided specific solutions to these equations for a number of initial density profiles, including a constant-density cloud (Section \ref{sec:const}), clouds with ``approximately constant'' densities (Section \ref{sec:appconst}), and polytropes (Section \ref{sec:polytropes}) -- including the case of an isothermal sphere ($\gamma = 1$). In these cases, we showed that the center of the cloud forms a point mass in a time of $\tau_{ff} = \sqrt{3\pi/(32{G}\bar\rho)}$, where $\bar\rho$ is the average density of the cloud, and was otherwise \emph{independent} of any other properties of the collapsing gas as long as the center of the cloud originally has a finite density. We also found that the density and velocity approached power-laws at this time, though the power-law index depended on the initial density profile of the cloud. After the singularity forms, the entire cloud (if it has a finite mass) accretes onto the central point mass in a small fraction of an infall time, as demonstrated by Figures \ref{fig:Aplots} and \ref{fig:gamma53}. We also calculated the accretion rates, which were highly time dependent (Figures \ref{fig:mdots} and \ref{fig:mdots_poly}) and were not necessarily finite at the time the core formed.

In Section \ref{sec:conditions} we showed analytically that our solutions form a singularity at the origin in precisely one freefall time, and we derived exact expressions for the scalings of the density and velocity near the point mass at this time, specifically given by Equations \eqref{rhoapp} and \eqref{vapp}; in these equations, $p$ is related to the initial mass profile of the cloud via $M_0(R) = R^3+CR^{p} +\ldots$ Agreement was shown with Penston's solution for self-similar, pressureless collapse ($\rho \propto R^{-12/7}$, $v \propto R^{1/7}$) only if $p = 5$, valid for polytropes, and otherwise we recovered the scalings derived by \citet{lyn88}. We also calculated the accretion rates onto the point mass, and we showed that, at the time the singularity formed, the accretion rate was not necessarily finite, and the asymptotic accretion rate for an isothermal sphere was shown to approach a constant value of $\dot{M} \simeq 2.7c_s^3/G$. 

We also analyzed the asymptotic ($t \gg \tau_{ff}$) scaling of the density, velocity, and accretion rate for clouds with $\rho \propto R^{-n}$ for $R \gg 1$ and $0 < n < 3$ (i.e., infinite mass clouds). It was found that $m(r,t) \rightarrow m(r)$, and that the density and velocity generally separated into power-laws of radius multiplied by power-laws of time. Interestingly, the only value of $n$ that preserved the temporal independence of the density was $n = 3/2$, with values less than (greater than) this value resulting in temporally decreasing (increasing) density profiles.

We showed that a turbulent pressure can be included in these models by setting ${\partial{p}}/{\partial{r}}/\rho \equiv (1-\beta^2)GM/r^2$, which asserts that the turbulent velocity scales as the local infall velocity. In this case, one recovers the same solutions for pressureless collapse, but with the velocities and variables replaced, respectively, by $v \rightarrow \sqrt{\beta}\,v$ and $\xi \rightarrow \xi/\sqrt{\beta}$. These solutions can be further generalized by allowing $\beta$ to be a function of $\xi$; while in principle this function could be completely arbitrary, one could use an energy equation similar to that of \citet{mur15}, which accounts for the generation of turbulent motions from gravitational potential energy, to close the system and generate a self-consistent set of solutions.

Finally, the temporal dependence of density perturbations -- which should be prevalent in any collapsing region given the importance of turbulence -- on top of an otherwise-smooth density profile was analyzed, the evolution of which should be nontrivial owing to the competition between tidal forces and self-gravity. We derived the inequality that differentiates between the shear and self-gravitating limits (Equations \ref{mcriter} or \ref{rhocriter}), and we considered a specific example with overdensities that were shear-dominated, on the cusp of self-gravitating, and dominated by self-gravity. When such overdensities are self-gravitating, they collapse under their own self-gravity to form infinitely thin shells on a timescale approximately given by Equation \eqref{taush}, those shells continuing their infall onto the point mass at the origin and generating an increase in the accretion rate (Figure \ref{fig:mdotcaustic}). 

As we noted above, our solutions for the collapse reproduce Penston's scaling at the moment the singularity forms (and conforms to the predictions of \citealt{lyn88} for flatter initial mass profiles). However, the amount of time over which the density and velocity sastisfy that scaling is exceedingly small: from the middle and right panels of Figure \ref{fig:gamma53}, the power-law indices for the density and velocity deviate significantly from $-12/7$ and $1/7$, respectively, on a timescale less than $0.1\tau_{ff}$. Thus, even though Penston's self-similar solution does manifest itself at the moment the point mass forms, it is not maintained, and the vast majority of the evolution of the infall proceeds non-self-similarly.

Our treatment here was highly idealized concerning our set of assumptions and, among those assumptions, we opted to neglect the influence of angular momentum and magnetic fields on the collapse. In the limit that they can be regarded as small in terms of their contribution to the radial momentum equation, these quantities can be evolved in a manner that is independent of the infall velocity and the density profile, each governed by the azimuthal (or poloidal) momentum equation (Equation \ref{azmom}) and the induction equation (Equation \ref{induction}). Surprisingly, one can find exact solutions for both the rotational velocity (Equation \ref{uex}) and each component of the magnetic field (Equations \ref{bperp} and \ref{bpar}), valid for \emph{any} initial angular momentum profile or field configuration. Our results show that, for a cloud that is initially isothermal and rotating rigidly, the centrifugal barrier becomes important at a radius of $R_{disk} \simeq (t_{rot}/t_{ff})^{7/2}$, where $t_{rot}$ is the original rotational period of the cloud; thus, for $t_{rot} \ll t_{ff}$, then this radius -- which likely corresponds to the location at which a rotationally-supported disk forms -- is on very small scales. We find that, if the magnetic field is initially uniform in the $z$-direction -- a case that has been considered in the literature (e.g., \citealt{pri12}) -- then the field lines become increasingly radial as the point mass forms, and, if the rotational velocity of the cloud is incorporated, toroidal fields are self-consistently generated from azimuthally-sheared radial and poloidal fields. These results may provide analytic estimates for the formation, collimation, and mass loss rates of protostellar jets from the first core \citep{bat14}. 

\acknowledgements
Support for this work was provided by NASA through the Einstein Fellowship Program, grant PF6-170150. I thank Phil Armitage, Jordan Mirocha, Chris McKee, Norm Murray, and Eliot Quataert for useful discussions and suggestions. {}{I also thank the anonymous referee for useful comments and suggestions.}

\appendix
\section{Rotation}
\label{sec:rotation}
Using the formalism developed here, one can analyze the effects of rotation when the azimuthal or poloidal velocity remains small in comparison to the radial velocity. At first this assumption might seem like a contradiction, as the initial condition for the collapsing cloud is $v_r(R,0) \equiv0$. However, if we return to the radial momentum equation, it is actually the \emph{temporal derivative} of $v_r$ that must exceed the centrifugal terms. In particular, we require $\partial{v_r}/\partial{t} \gtrsim v_\phi^2/r$; using our form for the radial velocity \eqref{vss}, this inequality becomes, as we expect, $v_\phi\lesssim \sqrt{GM/r}$.

Under the assumption that the azimuthal and poloidal velocities remain small relative to the infall velocity, the solutions for $v_r(r,t)$ and $M(r,t)$ remain unchanged from Equations \eqref{vss} and \eqref{Mofxi}. We can further neglect the nonlinear contributions to the azimuthal and poloidal momentum equations, and these equations then become

\begin{equation}
\frac{\partial{u}}{\partial{t}}+v_r\frac{\partial{u}}{\partial{r}}+\frac{v_ru}{r} = 0, \label{azmom}
\end{equation} 
where $u$ is either $v_\phi$ or $v_\theta$. This equation shows that the specific angular momentum, $\ell = u\,r$, is conserved along flow lines, and so the general solution for $u(r,t)$ is

\begin{equation}
u = \frac{1}{1-f^2}U_0\left(\frac{r}{1-f^2}\right), \label{uex}
\end{equation}
where $U_0$ is the initial rotational velocity of the cloud. As we expect, an initial rotational velocity of exactly zero yields zero rotation for all times.

A simple example, which has been analyzed numerically \citep{bat98}, is that of rigid rotation at a small angular velocity $\Omega$. We cannot assume this form for all $r$, as the cloud would be rotating infinitely fast (and  would violate our assumption of the smallness of $v_\phi$ in comparison to $v_r$) as we proceed farther from the center of the cloud. To maintain rigid rotation for all $r \lesssim R$ and small velocities at large $r$, we will therefore let

\begin{equation}
U_0(r) = \frac{\Omega{r}}{1+\frac{r^2}{R_C^2}}, \label{U0}
\end{equation}
which satisfies $u < v_r$ for all $r$ if $\Omega$ is sufficiently small. From equation \eqref{uex}, we immediately have that the time-dependent solution for this initial velocity profile is

\begin{equation}
u = \Omega{R_C}\frac{R}{(1-f^2)^2}\left(1+\frac{R^2}{(1-f^2)^2}\right)^{-2}. \label{uofxi}
\end{equation}

\begin{figure}[htbp] 
   \centering
   \includegraphics[width=0.47\textwidth]{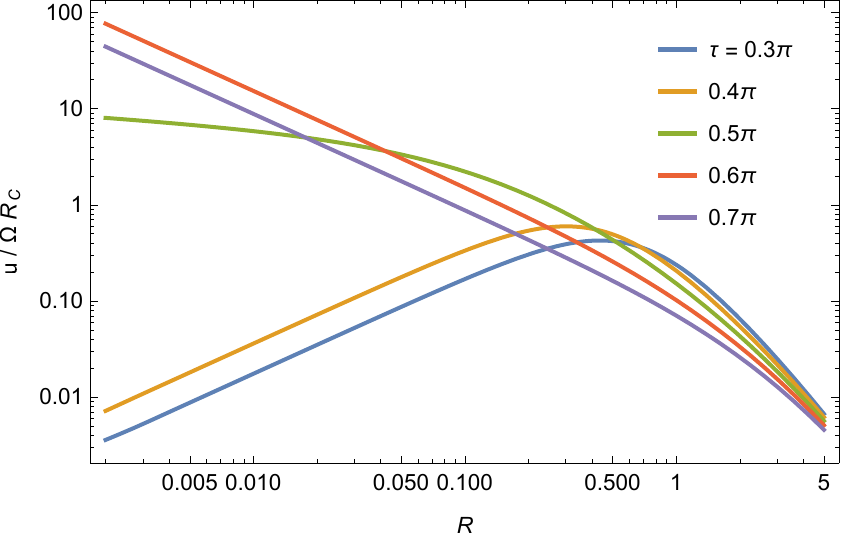} 
   \caption{The ratio $u/(\Omega{R_C})$ as a function of $R$ for an isothermal sphere at the times indicated in the legend. At $\tau = \pi/2$, when the point mass forms, the scaling of the rotational velocity is $u \propto R^{-1/7}$, in contrast to the scaling of the infall velocity $v \propto R^{1/7}$.}
   \label{fig:uofR}
\end{figure}

Figure \ref{fig:uofR} illustrates $u/(\Omega{R_C})$ as a function of $R$ for an isothermal sphere, and the different curves correspond to the different times indicated in the legend. We see that, at the time the singularity forms, the rotational velocity approaches a power-law near the origin; we can show, by using the asymptotic scaling of $f$ and $\xi$ for $\xi \gg 1$ (which corresponds to $R \ll 1$) in Equation \eqref{uofxi}, that this power-law should be $u \propto R^{-1/7}$, and Figure \ref{fig:uofR} confirms this prediction. For times $\tau > \pi/2$, the rotational velocity satisfies $u \propto R^{-1}$. 

Our treatment here assumed that the rotational velocity was dynamically unimportant, allowing us to neglect its contribution to the radial momentum equation. However, we see from Figure \ref{fig:uofR} that, once the core forms, the power-law index of $u$ becomes slightly negative, which contrasts the scaling of the radial velocity that, for a polytrope, satisfies $v_r \propto R^{1/7}$. Therefore, there will be a location within the flow at which the centrifugal term will become comparable to the advective term in the radial momentum equation, and this will occur when $v \simeq u$. Including all the relevant constants, the radius $R_{disk}$ at which $u = v$ for an isothermal sphere is

\begin{equation}
R_{disk}^{2/7} = \left(\frac{40}{3\pi}\right)^{1/7}\sqrt{\frac{3}{8\pi}}\frac{\Omega}{\sqrt{G\rho_C}} \simeq \frac{t_{ff}}{t_{rot}}, \label{rdisk}
\end{equation}
where $t_{ff} \simeq \sqrt{3\pi/(32G\rho_C)}$ is the freefall timescale and $t_{rot} = 2\pi/\Omega$ is the rotational period. This equation shows that $R_{disk} \simeq (t_{ff}/\tau_{rot})^{7/2}$, and is thus a very sensitive function of the ratio of the collapse timescale to the rotational period of the original cloud: even for a moderately small ratio of $0.1$, the scale at which one expects a disk to form is $\sim 10^{-4}$ the initial scale length of the cloud. Equation \eqref{rdisk} thus provides a good length scale required for numerical methods to resolve the inner, centrifugally-supported disks often recovered in simulations of protostellar collapse (e.g., \citealt{bat98, pri12, bat14}). 

\section{Magnetic Fields}
\label{sec:magnetic}
The role of magnetic fields in governing the behavior of a collapsing cloud has been intensely studied (e.g., \citealt{lyn03, mac06, ban06, pri12, bat14, lew15}), and analyses suggest that the field should be highly coupled to the gas in the early stages of infall \citep{mes56, nak02}. Thus, a reasonable assumption is that the equations of ideal MHD hold, in which case the evolution of the magnetic field is governed by the induction equation:

\begin{equation}
\frac{\partial\mathbf{B}}{\partial{t}}-\mathbf{\nabla}\times\left(\mathbf{v}\times\mathbf{B}\right) = 0. \label{induction}
\end{equation}
Using standard vector identities and the divergenceless nature of the magnetic field, this equation becomes

\begin{equation}
\frac{\partial\mathbf{B}}{\partial{t}}+(\mathbf{v}\cdot\nabla)\mathbf{B} = -\mathbf{B}\left(\nabla\cdot\mathbf{v}\right)+\left(\mathbf{B}\cdot\nabla\right)\mathbf{v}. \label{induction}
\end{equation}
In general, we must couple the induction equation to the momentum equation, as the magnetic pressure and tension can alter the infall of the gas. However, if we assume that magnetic forces remain subdominant to gravitational ones, so that $B^2/\rho \lesssim GM/r$ (equivalently, that the infall speed exceeds the Alfv\'en speed), then we can ignore the contribution of the magnetic field to the evolution of the gas. In this case, then, the velocity is still given by Equation \eqref{vss}, and we can solve Equation \eqref{induction} independently of the momentum equation. 

Before turning to this analysis, we point out that the vectors appearing in Equation \eqref{induction} are comprised of the \emph{coordinate} components (raised indices) of the fields, which are related to the \emph{physical} components (lowered indices) via $(v,B)_{r} = (v,B)^{r}$, $(v,B)_{\theta} = r(v,B)^{\theta}$, and $(v,B)_{\phi} = r\sin\theta(v,B)^{\phi}$ (e.g., \citealt{mih84}). From these transformations it is apparent that $v^{\theta}$ and $v^{\phi}$ are angular velocities (and a similar notion holds for $B^{\theta}$ and $B^{\phi}$), so we must be careful to use the appropriate representation. 

With this caveat in mind, first consider the $\theta$ and $\phi$ components of the magnetic field, the physical components of which we will collectively denote $\mathbf{B}_{\perp}$. Expanding out the derivatives and using the fact that the only non-zero velocity is $v_r$, we find that these components of the magnetic field satisfy

\begin{equation}
\frac{\partial{B_{\perp}}}{\partial{t}}+\frac{1}{r}\frac{\partial}{\partial{r}}\left(rv_rB_{\perp}\right) = 0.
\end{equation}
Here $B_{\perp}$ refers to either $B_{\phi}$ or $B_{\theta}$. We find that the general solution to this equation is

\begin{equation}
B_{\perp} = \frac{1}{r}\frac{\partial}{\partial{r}}\left[F_{\perp,0}\left(\frac{r}{1-f^2}\right)\right], 
\end{equation}
where $F_{\perp,0}$ is the flux of the perpendicular field. To relate the function $F_{\perp,0}$ to the field itself, we note the following useful identity, which can be proved relatively easily using the chain rule:

\begin{equation}
\frac{\partial}{\partial{r}}\left(\frac{r}{1-f^2}\right) = \frac{2+f\xi\frac{\partial{\ln{}M}}{\partial\ln{r}}}{\left(1-f^2\right)\left(2+3f\xi\right)}.
\end{equation}
It then follows that the perpendicular component of the field is given by

\begin{equation}
B_{\perp} = \frac{2+f\xi\frac{\partial{\ln{}M}}{\partial\ln{r}}}{\left(1-f^2\right)^2\left(2+3f\xi\right)}B_{\perp,0}\left(\frac{r}{1-f^2}\right). \label{bperp}
\end{equation}

For the radial component of the magnetic field, which we will denote $B_{r}$, expanding out the right-hand side of Equation \eqref{induction} gives

\begin{equation}
\frac{\partial{B_{r}}}{\partial{t}}+v\frac{\partial{B_{r}}}{\partial{r}} = -\frac{2B_{r}v}{r},
\end{equation}
and some simple rearrangements of this equation then yields

\begin{equation}
\frac{\partial}{\partial{t}}\left(r^2B_{r}\right)+v_r\frac{\partial}{\partial{r}}\left(r^2B_{r}\right) = 0.
\end{equation}
Recalling the fact that the two-sphere has an area $A \propto r^2$, this equation just reflects flux conservation in the radial direction. Comparing this to the continuity equation then gives as the general solution

\begin{equation}
B_{r} = \left(1-f^2\right)^{-2}B_{r,0}\left(\frac{r}{1-f^2}\right), \label{bpar}
\end{equation}
where $B_{r,0}$ is the initial radial component of the magnetic field.

As an example, consider the case where the field is initially uniform in the $z$-direction:

\begin{equation}
\mathbf{B}_{\perp,0} = B_{\theta} = -B_0\sin\theta, \label{bth0}
\end{equation}
\begin{equation}
B_{r,0} = B_0\cos\theta, \label{br0}
\end{equation}
where $B_0$ is the initial strength of the field; note that the inclusion of an angular dependence on $\mathbf{B}$ does not alter our results. Equations \eqref{bperp} and \eqref{bpar} then give

\begin{equation}
B_{\theta} = -\frac{2+f\xi\frac{\partial{\ln{}M}}{\partial\ln{r}}}{\left(1-f^2\right)^2\left(2+3f\xi\right)}B_0\sin\theta, \label{bth}
\end{equation}
\begin{equation}
B_r = \frac{B_0\cos\theta}{\left(1-f^2\right)^{2}}. \label{br}
\end{equation}

\begin{figure}[htbp] 
   \centering
   \includegraphics[width=.47\textwidth]{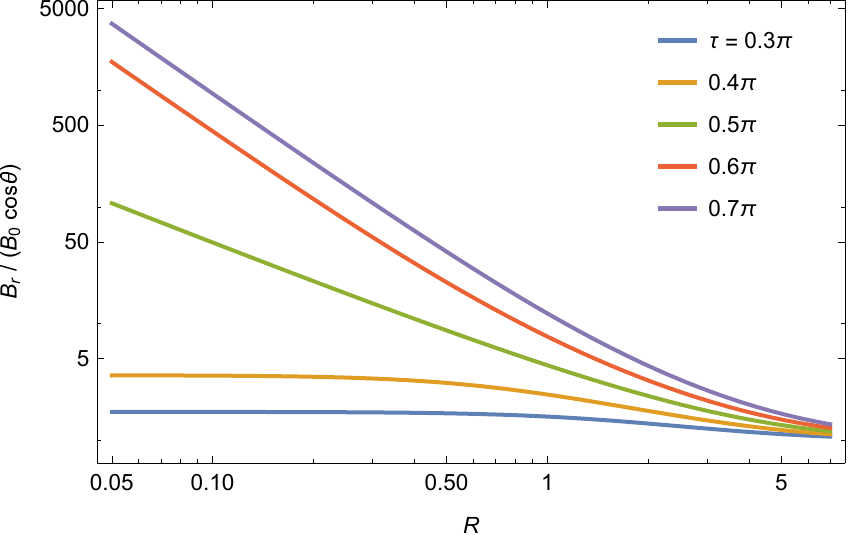} 
   \hspace{0.02\textwidth}
   \includegraphics[width=0.47\textwidth]{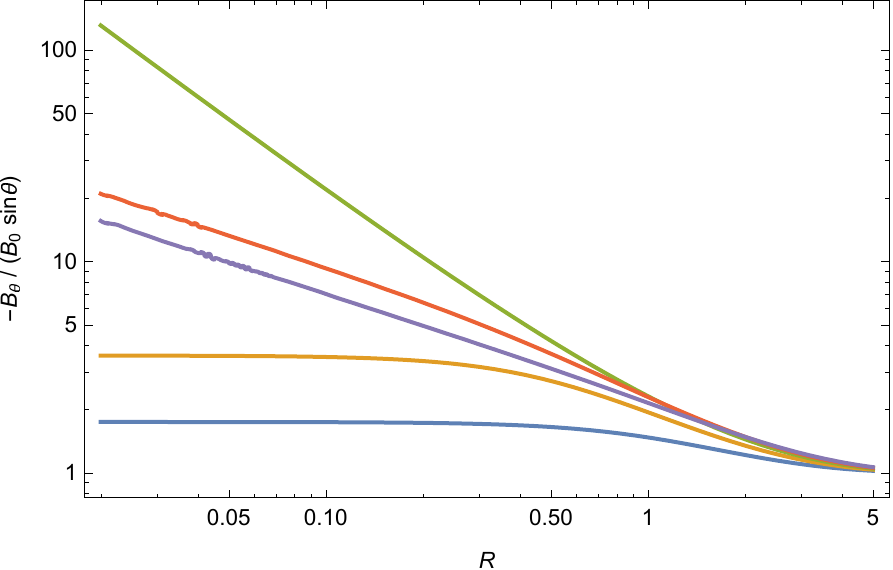}
   \caption{The radial component of the field (left panel) and the poloidal component (right panel) when the density profile of the collapsing cloud is appropriate to an isothermal sphere, where the different curves correspond to the different times indicated by the legend in the left-hand panel. }
   \label{fig:BofR}
\end{figure}

Figure \ref{fig:BofR} shows the radial component of the field (left panel) and the poloidal component (right panel) when the initial density profile of the cloud is appropriate to an isothermal sphere. The different lines correspond to the times in the legend. At $\tau = \pi/2$, the radial and poloidal components of the field approach power laws with the same, negative power-law index. Using Equations \eqref{mapp} and \eqref{rhoapp} in Equations \eqref{bth} and \eqref{br}, we find $B_r \propto B_{\theta} \propto R^{-8/7}$ for an isothermal sphere, and 

\begin{equation}
B_r \propto B_{\theta} \propto R^{\frac{12-4p}{2p-3}} \propto \rho^{2/3}, \label{Brapp}
\end{equation}
for arbitrary initial mass profiles of the form $M_0(R) = R^3+\mathcal{O}(R^{p})$. For times $\tau > \pi/2$, we find $B_r \propto r^{-2}$ and $B_{\theta} \propto r^{-1/2}$. 

\begin{figure}[htbp] 
   \centering
   \includegraphics[width=.325\textwidth]{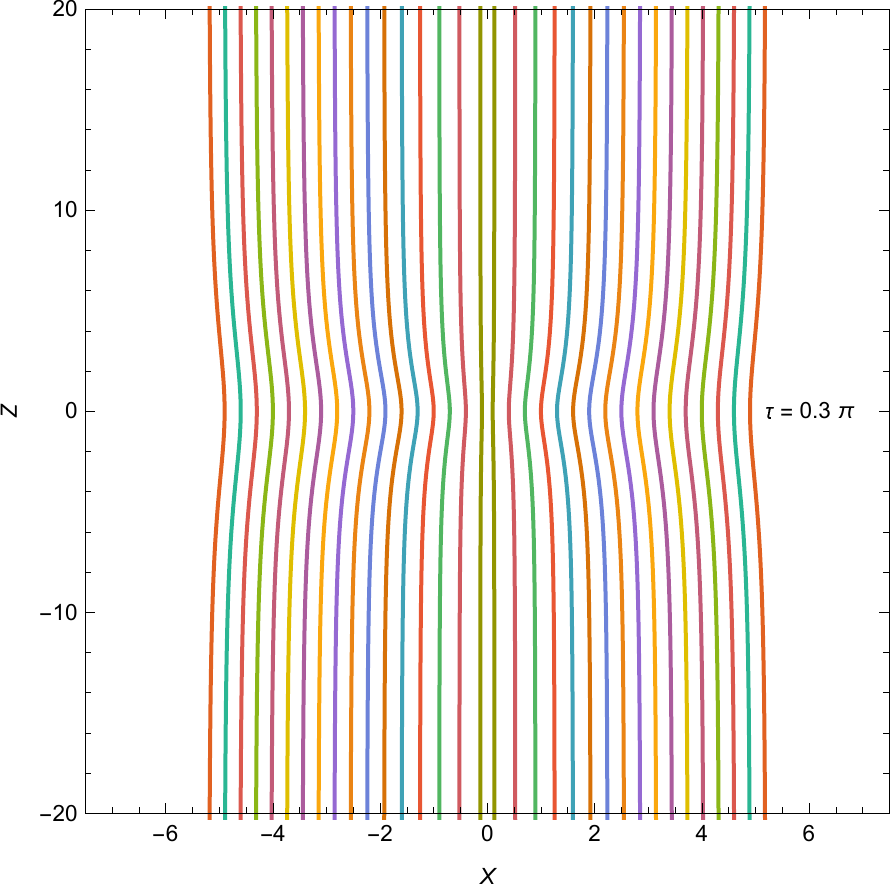} 
   \includegraphics[width=.325\textwidth]{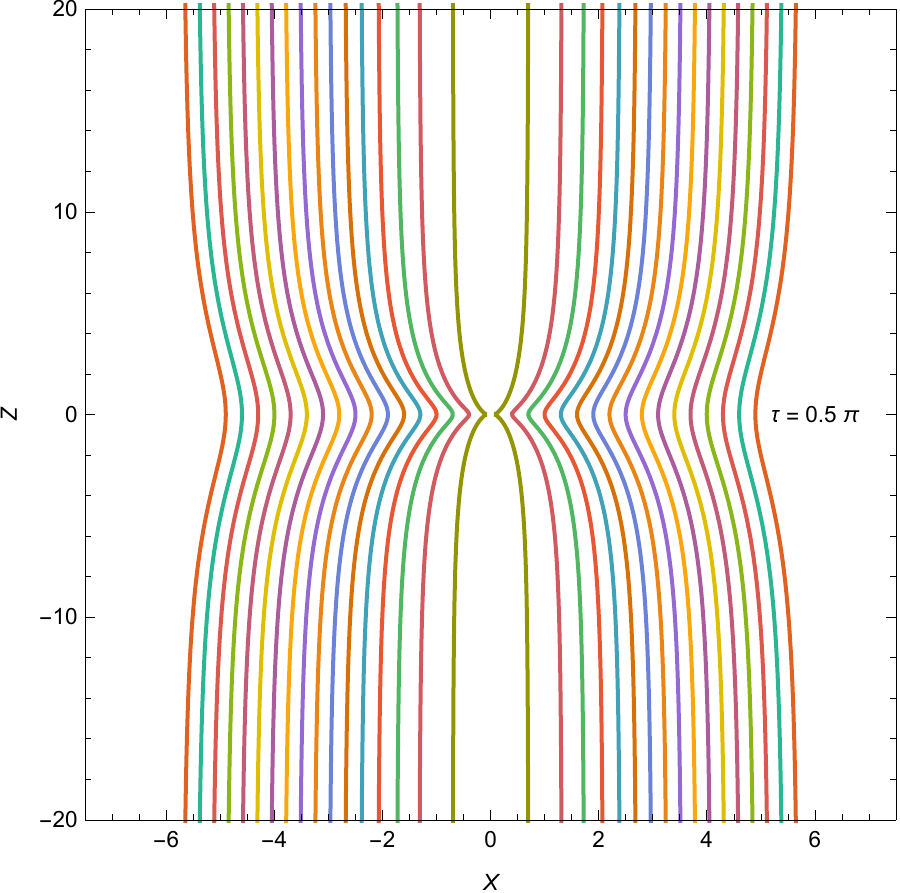}
   \includegraphics[width=.325\textwidth]{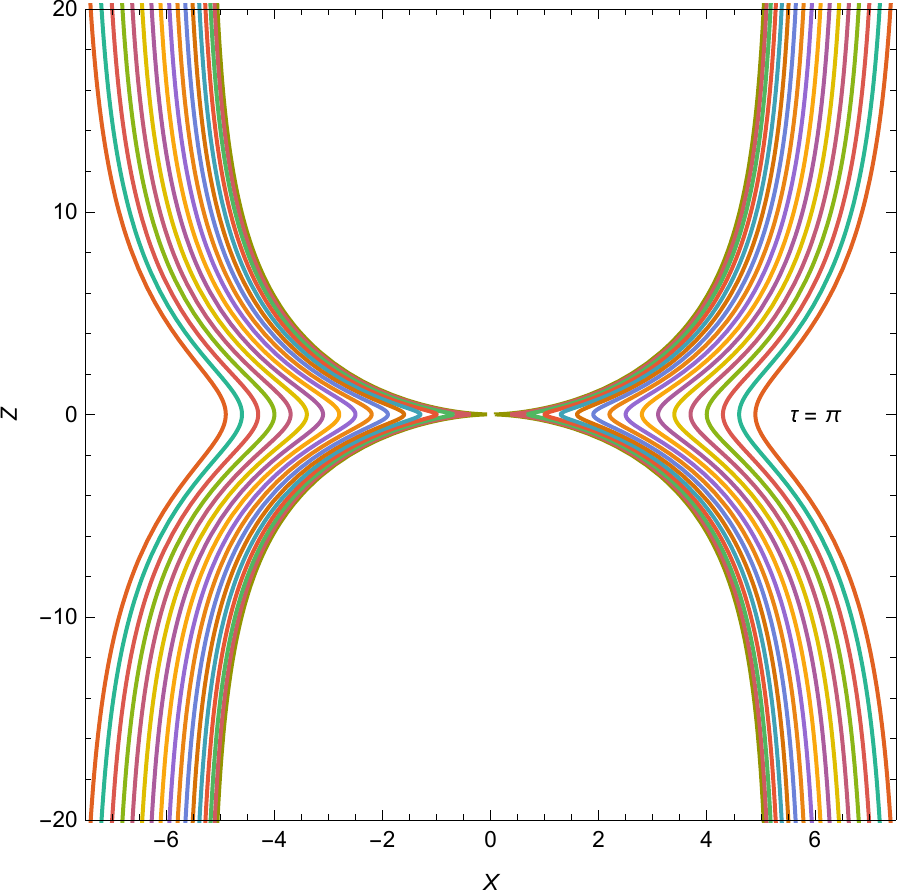}
   \caption{The magnetic field lines for a field configuration that initially satisfies $\mathbf{B} = B_0\hat{z}$ at three different times: $\tau = \pi/2$ (left-hand panel), $\tau = \pi$ (middle panel), and $\tau = 2\pi$ (right-hand panel).  }
   \label{fig:fieldlines}
\end{figure}

Figure \ref{fig:fieldlines} gives the field lines at three different times -- prior to the formation of the point mass (left-hand panel), when the singularity forms (middle panel), and after the singularity forms (right-hand panel) -- for the initial conditions $\mathbf{B} = B_0\hat{z}$ and an isothermal sphere. As more mass falls toward the origin, the field lines become increasingly radial, which is consistent with the notion that the gas ``drags'' the field lines with it. The middle panel of this figure agrees quite well with numerical findings (compare the middle panel of our Figure \ref{fig:fieldlines} to Figure 1 of \citealt{bat14}). 

Maintaining the exact constants in Equations \eqref{bth} and \eqref{br}, we find that, for $R \lesssim 1$, the field lines at the time the singularity forms satisfy

\begin{equation}
R(\theta) = R_0\left(\sin\theta\right)^{-\frac{2p-3}{3}},
\end{equation}
where $R_0$ is the location of the field line at $\theta = \pi/2$. Because the field is symmetric about the midplane, the lines are exactly vertical at $\theta = 0$. However, it is apparent from the middle panel of Figure \ref{fig:fieldlines} that for $\theta \gtrsim 0$, the field lines make a smaller angle with the $X$-$Z$ plane that would intersect a disk of finite thickness $H$. For small $R$ the angle made by these lines is less than $60^{\circ}$, meaning that the configuration should be capable of launching a magnetocentrifugally-driven wind \citep{bla82} consistent with those seen in simulations \citep{pri12, bat14}.

Finally, there is no difficulty in adding an azimuthal component to the magnetic field; however, we see from Equation \eqref{bperp} that we must embed within the cloud an initial $B_{0,\phi}$ if the evolution of the field is to be nontrivial. Starting with such an artificial field is somewhat undesirable from a physical standpoint, as the azimuthal component of the field should be naturally generated by the rotation inherent in the collapsing material. 

Surprisingly, we can still make a large amount of analytic progress in the case where we include the effects of azimuthal rotation on the generation of the field. As we did in Appendix \ref{sec:rotation}, if we let the azimuthal velocity be denoted by $u(r)$, then we can write the generalized version of Equation \eqref{induction} as

\begin{equation}
\frac{\partial}{\partial{t}}\left(rB_{\phi}\right)+\frac{\partial}{\partial{r}}\left(rv_rB_{\phi}\right) = r^2B_r\frac{\partial}{\partial{r}}\left(\frac{u}{r}\right) +\sin\theta{B_\theta}\frac{\partial}{\partial\theta}\left(\frac{u}{\sin\theta}\right). \label{Bphigen}
\end{equation}
Using the fact that $u$ satisfies Equation \eqref{azmom}, we can show that the general solution to this equation is

\begin{equation}
B_{\phi} = \frac{2+f\xi\frac{\partial{\ln{}M}}{\partial\ln{r}}}{\left(1-f^2\right)^2\left(2+3f\xi\right)}B_{\phi,0}\left(\frac{r}{1-f^2}\right)+2rB_r\frac{\partial}{\partial{r}}\left(\frac{u}{v_r}f^2\right)-2B_{\theta}\cot\theta\frac{u}{v_r}f^2. \label{bphitot}
\end{equation}
The first term is just the homogeneous solution, and the second two terms can be considered particular solutions that are amplified by the azimuthal velocity.

At face value, Equation \eqref{bphitot} is somewhat complicated and difficult to interpret. However, the fundamental nature of the terms proportional to $B_r$ and $B_{\theta}$ -- which are the two terms driving the field generation within the flow -- can be understood as follows: consider a radial field line connecting two fluid elements within the collapsing cloud. As the cloud starts to rotate, the two fluid elements will generally be sheared apart, and since the field lines are frozen in, this shearing will generate a non-zero toroidal field; in particular, if the fluid is rotating faster at smaller $r$, so $\partial\Omega/\partial{r} > 0$, the shear will generate a positive $B_{\phi}$, while $\partial\Omega/\partial{}r < 0$ implies a negative $B_{\phi}$. Only $\partial\Omega/\partial{}r \equiv 0$ yields no growth of $B_{\phi}$, which, recalling that $\Omega = u/r$, is reflected by the first term on the right-hand side of Equation \eqref{Bphigen}. 

Now envision a poloidal field line connecting two neighboring fluid elements. Along this line the two elements are separated by an amount $\Delta{s}$ in \emph{cylindrical} radius, and differential rotation along the $z$-axis will then shear out the field line. In particular, if $u/(r\sin\theta)$ is increasing with $\theta$, the shear will generate a positive toroidal field, and a negative toroidal field is created when $u/(r\sin\theta)$ decreases with increasing $\theta$. The only way to prevent the generation of a toroidal field from the poloidal component is to have constant angular velocity on cylinders, which is confirmed by the second term on the right-hand side of Equation \eqref{Bphigen}.

\begin{figure}[htbp] 
   \centering
   \includegraphics[width=0.5\textwidth]{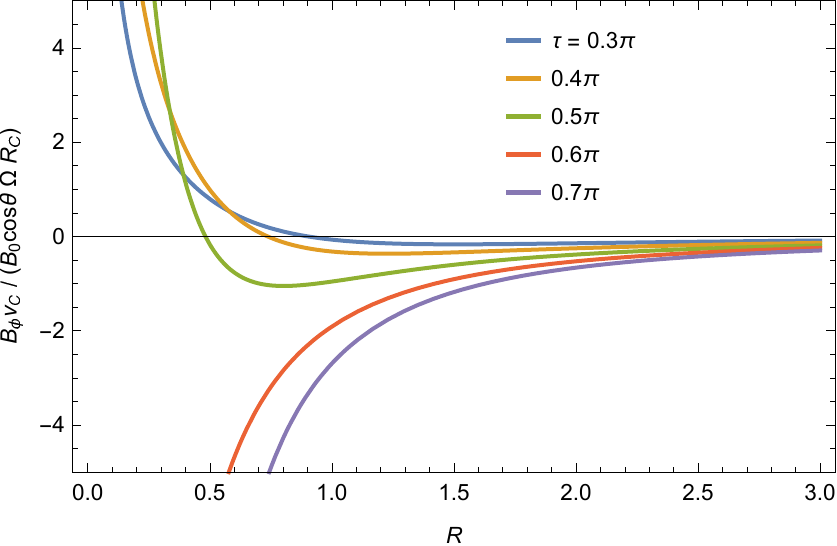}  
  \caption{The toroidal component of the magnetic field, suitably normalized, when the cloud initially has an isothermal density profile, the cloud is initially rotating as $U_0 = \Omega{R}/(1+R^2)$, $B_r$ is appropriate to the case of a field that is initially uniform in $z$, and the initial $B_{\phi,0}$ is zero (i.e., the field is entirely generated from the shear in the flow).}
   \label{fig:bphi}
\end{figure}

Figure \ref{fig:bphi} shows the (appropriately normalized) $\phi$-component of the field for an initially-isothermal mass profile at the times shown in the legend. Here the cloud is initially rotating with the azimuthal velocity considered in Appendix \ref{sec:rotation}, $U_0 = \Omega{R}/(1+R^2)$, and the magnetic field is initially uniform in the $z$-direction; there is thus no $B_{\phi,0}$ in this case, and the toroidal field is only generated by the shear in the flow. It is apparent that, when $R \lesssim 1$, the induced toroidal field is in the direction of the angular velocity; however, for $R \gtrsim 1$, the field lines change sign.

The reason for this behavior is due to the fact that both $B_r$ and $B_{\theta}$ are being sheared to create $B_{\phi}$, but the sign of the toroidal field generated by $B_r$ differs from that generated by $B_\theta$. In particular, the decreasing $u/(R\sin\theta)$ makes the shear in the $\theta$-direction more dominant at $R \ll 1$, which leads to a positive $B_{\phi}$ (recall that the sign of $B_{\theta}$ here is negative), while the negative $\partial\Omega/\partial{R}$ at large $R$ more efficiently generates a negative toroidal field from $B_r$. In this case, it is actually straightforward to determine the initial ($0 \le \tau \le \pi/2$) growth rate for $B_{\phi}$ from Equation \eqref{bphitot}, and we find

\begin{equation}
B_{\phi} \simeq B_0t\cos\theta\left\{r\frac{\partial\Omega}{\partial{r}}+\Omega\right\} = B_0{t}\cos\theta\frac{\partial}{\partial{r}}\left[\Omega{r}\right] = B_0t\cos\theta\frac{\partial{u}}{\partial{r}}.
\end{equation}
Thus, for the specific initial field configuration of $\mathbf{B} = B_0\mathbf{z}$, the toroidal field growth rate is proportional to the gradient in the physical velocity and not the angular velocity.

\begin{figure}[htbp] 
   \centering
      {\includegraphics[width=0.325\textwidth]{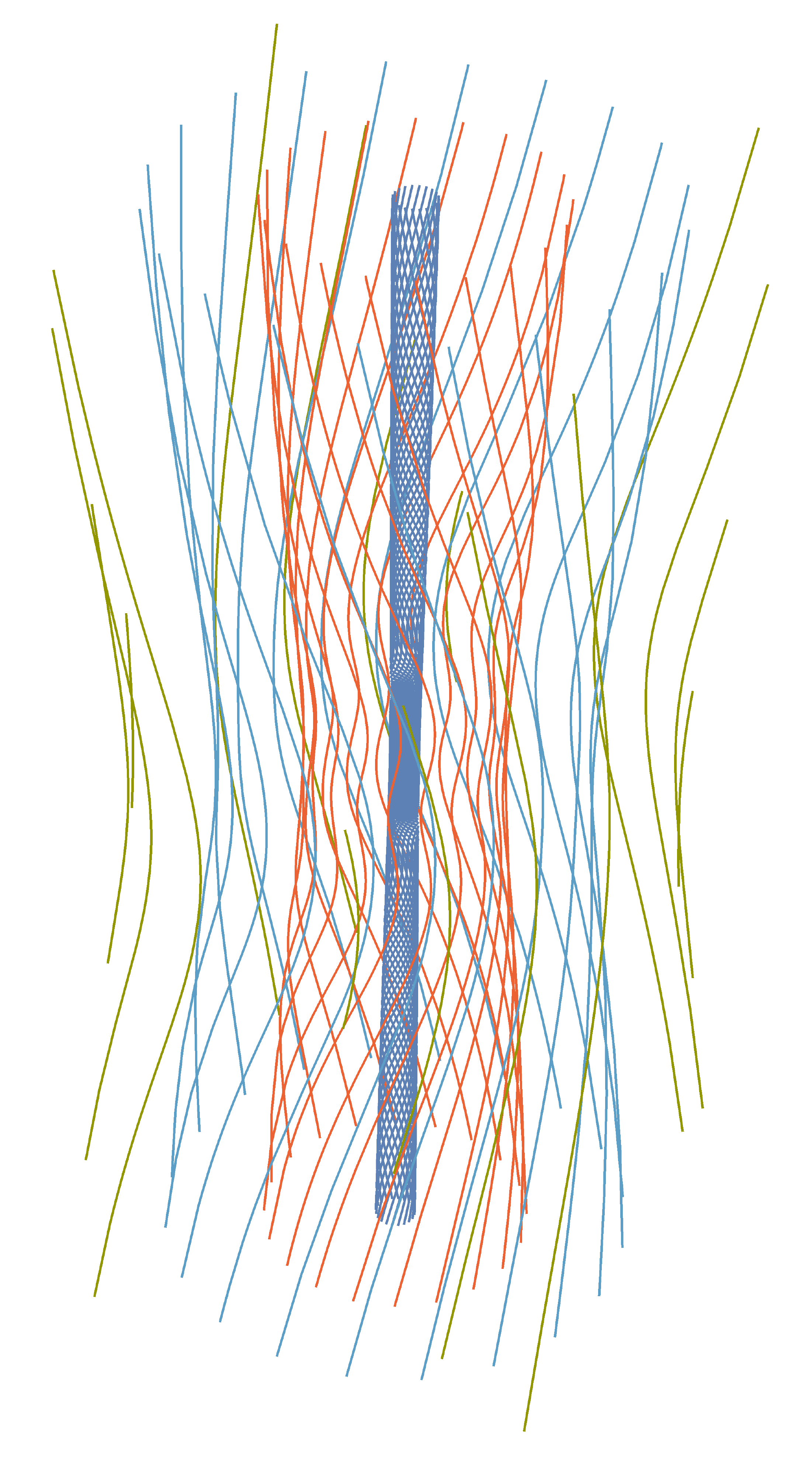}} 
  {\includegraphics[width=0.325\textwidth]{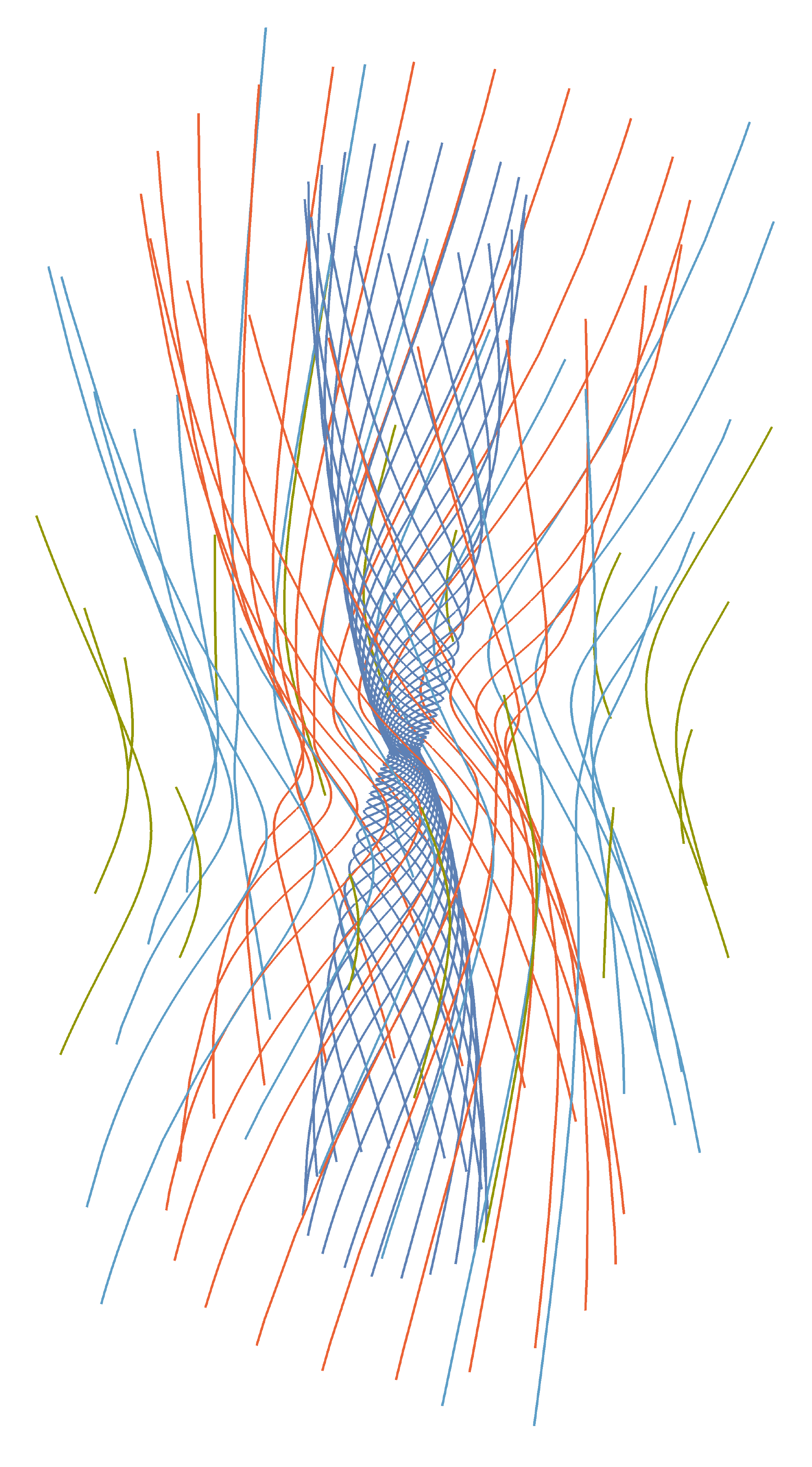}}
   {\includegraphics[width=0.325\textwidth]{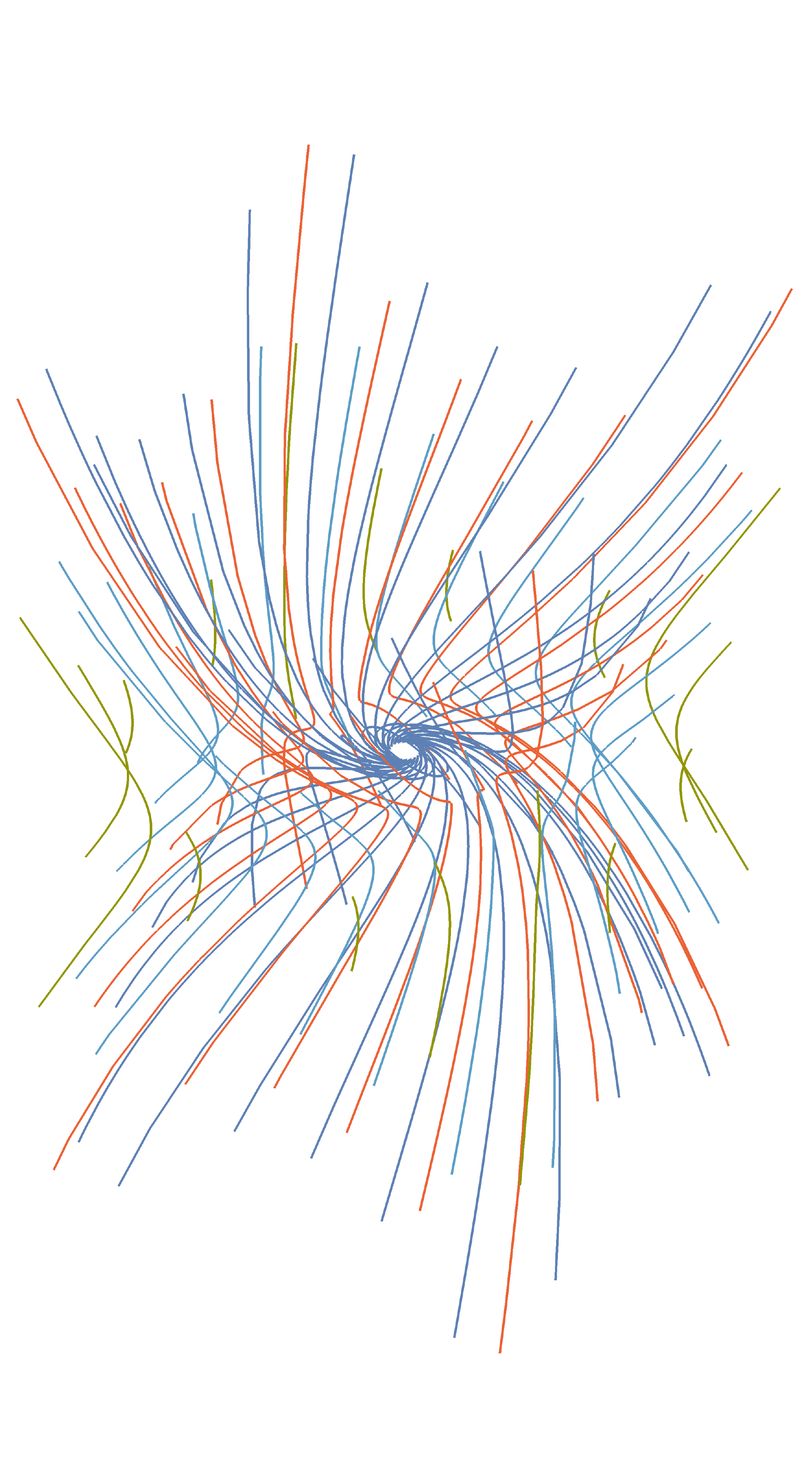}} 
   \caption{The field lines at $\tau = 0.4\pi$ (left panel), $\tau = \pi/2$ (middle panel) and $\tau = 0.6\pi$ (right panel) for a cloud that is initially isothermal, has $U_0$ given by Equation \eqref{U0}, and has $\mathbf{B}_0 = B_0\mathbf{z}$. Here we set $\Omega{R_C}/v_C = 0.1$, and the plot range extends from -1 to 1 in $X$ and $Y$ (directions perpendicular to the rotation axis) and from $-2$ to 2 in $Z$ (along the rotation axis). Movies showing how the field lines evolve in time can be found \href{http://w.astro.berkeley.edu/~eric_coughlin/movies.html}{here}.}
   \label{fig:maglines}
\end{figure} 

Figure \ref{fig:maglines} shows the magnetic field lines at $\tau = 0.4$ (just before the singularity forms; left panel), $\tau = 0.5$ (when the singularity forms; middle panel) and $\tau = 0.6\pi$ (just after the singularity forms; right panel) for the same configuration in Figure \ref{fig:fieldlines}. In this case we chose $\Omega{R_C}/v_C = 0.1$, and the axes extend from -1 to 1 in $X$ and $Y$ (perpendicular to the rotation axis) and from -2 to 2 in $Z$ (parallel to the rotation axis). This figure demonstrates that the lines are tightly wound in a hourglass-like shape about the $z$-axis when the point mass forms. A movie showing the evolution of the fieldlines can be found at \href{http://w.astro.berkeley.edu/~eric_coughlin/movies.html}{this website}.

\bibliographystyle{apj}
\bibliography{/Users/ecogs89/Library/texmf/bibtex/bib/mybib}

\end{document}